\documentclass[preprint,showpacs,preprintnumbers,amsmath,amssymb,12pt]{revtex4-1}
\usepackage{lipsum}
\usepackage{setspace}
\usepackage[colorlinks,linkcolor = black,anchorcolor = blue,urlcolor = blue,	citecolor = blue]{hyperref}
\usepackage{float}
\usepackage{graphicx}
\usepackage{makecell}
\usepackage{lineno}

\begin{document}
\title{Nodal $\textsl{s}_\pm$ Pairing Symmetry in an Iron-Based Superconductor with only Hole Pockets}

\author{Dingsong Wu$^{1,2,\dagger}$, Junjie Jia$^{1,2,\dagger}$, Jiangang Yang$^{1,2,\dagger}$, Wenshan Hong$^{1,2}$, Yingjie Shu$^{1,2}$, Taimin Miao$^{1,2}$, Hongtao Yan$^{1,2}$, Hongtao Rong$^{1,2}$, Ping Ai$^{1,2}$, Xing Zhang$^{1,2}$, Chaohui Yin$^{1,2}$, Chenlong Li$^{3}$, Shenjin Zhang$^{3}$, Fengfeng Zhang$^{3}$, Feng Yang$^{3}$, Zhimin Wang$^{3}$, Nan Zong$^{3}$, Lijuan Liu$^{3}$, Rukang Li$^{3}$, Xiaoyang Wang$^{3}$, Qinjun Peng$^{3}$, Hanqing Mao$^{1,2,5}$, Guodong Liu$^{1,2,5}$, Shiliang Li$^{1,2,5}$, Huiqian Luo$^{1,2,5}$, Xianxin Wu$^{4}$, Zuyan Xu$^{3}$, Lin Zhao$^{1,2,5\bigstar}$ and X. J. Zhou$^{1,2,5\bigstar}$}

\affiliation{
\\$^{1}$Beijing National Laboratory for Condensed Matter Physics, Institute of Physics, Chinese Academy of Sciences, Beijing 100190, China.
\\$^{2}$University of Chinese Academy of Sciences, Beijing 100049, China.
\\$^{3}$Technical Institute of Physics and Chemistry, Chinese Academy of Sciences, Beijing 100190, China.
\\$^{4}$Institute of Theoretical Physics, Chinese Academy of Sciences, Beijing 100190, China.
\\$^{5}$Songshan Lake Materials Laboratory, Dongguan, Guangdong 523808, China.
}

\date{\today}
\maketitle
\noindent{\bf The origin of the high temperature superconductivity in the iron-based superconductors remains elusive after being extensively studied for more than a decade. Determination of the pairing symmetry is essential in understanding the superconductivity mechanism. In the iron-based superconductors that have hole pockets around the Brillouin zone center and electron pockets around the zone corners, the pairing symmetry is generally considered to be \textbf{$\textsl{s}_\pm$}, endowing a sign change in the superconducting gap between the hole and electron pockets. For the iron-based superconductors with only hole pockets, however, a couple of pairing scenarios have been proposed but the exact symmetry is still highly controversial. Here we report our determination of the pairing symmetry in \bf{KFe$_2$As$_2$} which is a prototypical iron-based superconductor with hole pockets both around the zone center and around the zone corners. By taking laser-based angle resolved photoemission measurements with super-high resolution and at ultra-low temperature, we have precisely determined the superconducting gap distribution and identified the locations of the gap nodes on all the Fermi surface around the zone center and the zone corners. The complete superconducting gap structure, in combination with the observation of the spin resonance in neutron scattering, provides strong evidence on the \textbf{$\textsl{s}_\pm$} pairing symmetry in \bf{KFe$_2$As$_2$} with a gap sign reversal between the hole pockets around the zone center and the hole pockets around the zone corners. These results unify the pairing symmetry in the hole-doped iron-based superconductors and point to the spin fluctuation as the pairing glue in generating superconductivity.
}

\vspace{3mm}

The superconductivity mechanism of the iron-based superconductors, discovered more than a decade ago\cite{HHosono2008YKamihara, ZXZhao2008ZARen, DJohrendt2008MRotter, CQJin2008XCWang, MWu2008FHsu}, remains an outstanding issue in condensed matter physics\cite{IIMazin2011PJHirschfeld, DHLee2011FWang, HHosono2015, RFernandes2022}. The identification of the pairing symmetry and the revelation of the pairing medium are two key prerequisites in understanding the superconductivity mechanism. As a typical multiorbital system, a number of pairing symmetries have been proposed in iron-based superconductors that are associated with different pairing glues\cite{IIMazin2011PJHirschfeld, DHLee2011FWang, HHosono2015, RFernandes2022}. In most iron-based superconductors with the hole-like Fermi surface around the zone center and the electron-like Fermi surface around the zone corners, an $\textsl{s}_\pm$ pairing symmetry is expected from spin fluctuations based on weak coupling Fermi surface nesting picture\cite{IIMazin2008, HAoki2008KKuroki} or strong coupling local magnetic picture\cite{AElihu2008SQimiao, JPHu2008KSeo}. Alternatively, an $\textsl{s}_{++}$ pairing symmetry have also been proposed owing to the orbital fluctuations\cite{SOnari2010HKontani, YOno2010YYanagi2, YOno2010YYanagi}. In some iron-based superconductors with only electron-like Fermi surface\cite{DLFeng2011YZhang, HDing2011TQian, XJZhou2011DXMou, XJZhou2012DFLiu, XJZhou2016LZhao},  the $\textsl{d}$-wave pairing is found to be dominant from electron scatterings between the zone corners by spin fluctuations\cite{HAoki2008KKuroki, DJScalapino2011TAMaier, DHLee2011FWang2}. With increasing band hybridization, it was also shown that the system may undergo a transition from a $\textsl{d}$-wave state to an $\textsl{s}$-wave state where the gap changes sign between the hybridized pockets around the zone corners\cite{AVChubukov2012MKhodas}. For the iron-based superconductors with only hole-like Fermi surface, theoretical studies indicate that the $\textsl{s}_\pm$ and $\textsl{d}$ pairing channels are in close competition and the $\textsl{d}$-wave pairing is likely to excel at high hole doping\cite{BABernevig2011RThomale, AVChubukov2011SMaiti2, KKuroki2011KSuzuki}. A totally different pairing symmetry was also proposed that the sign reversal occurs between the two hole pockets around the zone center\cite{AVChubukov2012SMaiti, AVChubukov2017OVafek}. However, in the iron-based superconductors with only hole pockets, there is still a controversy about whether the pairing is
nodal\cite{HEisaki2009HFukazawa, SYLi2010JKDongPRL, TShibauchi2010KHashimoto, LTaillefer2012JPhReid, SShin2012KOkazaki, HDing2013NXu_BKFA, SLDrechsler2013MAbdel-Hafiez, YMatsuda2014DWatanabe, RProzorov2016KCho} or
nodeless\cite{MZolliker2011HKawano-Furukawa, CMeingast2016FHardy} and whether it is
$\textsl{d}$ wave\cite{SYLi2010JKDongPRL, TShibauchi2010KHashimoto, LTaillefer2012JPhReid, SLDrechsler2013MAbdel-Hafiez, LTaillefer2013FFTafti, RProzorov2014HKim} or
$\textsl{s}$ wave\cite{MZolliker2011HKawano-Furukawa, SShin2012KOkazaki, HDing2013NXu_BKFA, CMeingast2014FHardy2, YMatsuda2014DWatanabe, CMeingast2016FHardy, RProzorov2016KCho, HHKlauss2020VGrinenko, JZhao2020SDShen}.
Direct determination of the gap structure in the purely hole-doped iron-based superconductors is important to ascertain its pairing symmetry and establish a unified picture of the pairing mechanism in the iron-based superconductors.

As the end member of the hole-doped (Ba$_{1-x}$K$_x$)Fe$_2$As$_2$, KFe$_2$As$_2$ is a prototypical system for studying superconductivity mechanism in the purely hole-doped iron-based superconductors. It has unique electronic structure that consists of only hole pockets\cite{HDing2009TSato, SUji2013TTerashima, HHisatomo2014TYoshida}.  Even though it has the highest hole-doping level in iron-based superconductors, it is still superconducting with a T$_C$ of $\sim$3.5\,K\cite{DJohrendt2008MRotter, EHiroshi2010KKunihiro}. KFe$_2$As$_2$ is also one of the cleanest stoichiometric iron-based superconductors with a remarkably large residual resistivity rate\cite{EHiroshi2010KKunihiro} which is ideal for investigating its intrinsic electronic structure and physical properties. It has been found that KFe$_2$As$_2$ has a large Sommerfeld coefficient which behaves similarly to heavy Fermion materials\cite{UShinya2010TTaichi, XFWang2011JSKim, LBoeri2012MAbdel-Hafiez, SUji2013TTerashima, CMeingast2013FHardy, HVLohneysen2016FEilers, XHChen2016YPWu}, indicating the presence of strong electron correlation due to its proximity to the putative Mott insulating state\cite{MCapone2014Lde'Medici}. However, since KFe$_2$As$_2$ does not have a flat band from f-electrons near the Fermi level like the heavy Fermion materials\cite{KTMoore2009, CPfleiderer2009}, the electronic origin of its heavy-electron behaviors remains elusive. More importantly, although extensive studies have been carried out to probe the pairing symmetry of KFe$_2$As$_2$, the pairing nature has not been pinned down and remain controversial\cite{HEisaki2009HFukazawa, SYLi2010JKDongPRL, TShibauchi2010KHashimoto, LTaillefer2012JPhReid, SShin2012KOkazaki, SLDrechsler2013MAbdel-Hafiez, YMatsuda2014DWatanabe, RProzorov2016KCho, MZolliker2011HKawano-Furukawa, CMeingast2016FHardy, LTaillefer2013FFTafti, RProzorov2014HKim, CMeingast2014FHardy2, HHKlauss2020VGrinenko, JZhao2020SDShen}. In principle, angle resolved photoemission spectroscopy (ARPES) can directly determine the momentum-dependent gap distribution along the Fermi surface\cite{ZXShen2021}. However, the very low T$_C$ ($\sim$3.5\,K) and tiny gap size (0$\sim$1\,meV) in KFe$_2$As$_2$ make the ARPES measurements of its superconducting gap structure highly challenging. There has been only one report on ARPES measurement of the superconducting gap in KFe$_2$As$_2$\cite{SShin2012KOkazaki}, but only the momentum space around the Brillouin center was covered due to the limitation of the laser source. Moreover, the reliability of the reported results in ref. \onlinecite{SShin2012KOkazaki} is questionable because of the data quality and the gap analysis problems, as we will discuss later.

In order to determine the superconducting gap structure of KFe$_2$As$_2$ and other unconventional superconductors with low T$_C$ and small superconducting gap, we have developed a laser ARPES system with ultra-low temperature ($\sim$0.8\,K) and super-high energy resolution ($\textless$\,0.5\,meV) (see Methods). We also grew high quality KFe$_2$As$_2$ single crystals with a T$_C$ at 3.7\,K and a sharp superconducting transition ($\sim$0.2\,K) (see Methods and Fig. S1 in Supplementary Materials). We have clearly resolved the complete Fermi surface topology and band structures of KFe$_2$As$_2$, particularly those near the Brillouin zone corners by taking the advantage of surface reconstruction. We have precisely determined the momentum-dependent superconducting gap on each Fermi surface and identified the locations of the gap nodes. These results unravel the electronic origin of the huge mass enhancement and strong electron correlation in KFe$_2$As$_2$. They also provide strong evidence on the $\textsl{s}_\pm$ pairing symmetry with a sign reversal between the zone center and zone corners in the system with only hole pockets and point to the spin fluctuation as the pairing glue of the hole-doped iron-based superconductors.

\begin{figure*}[tbp]
\begin{center}
\includegraphics[width=1\textwidth,angle=0]{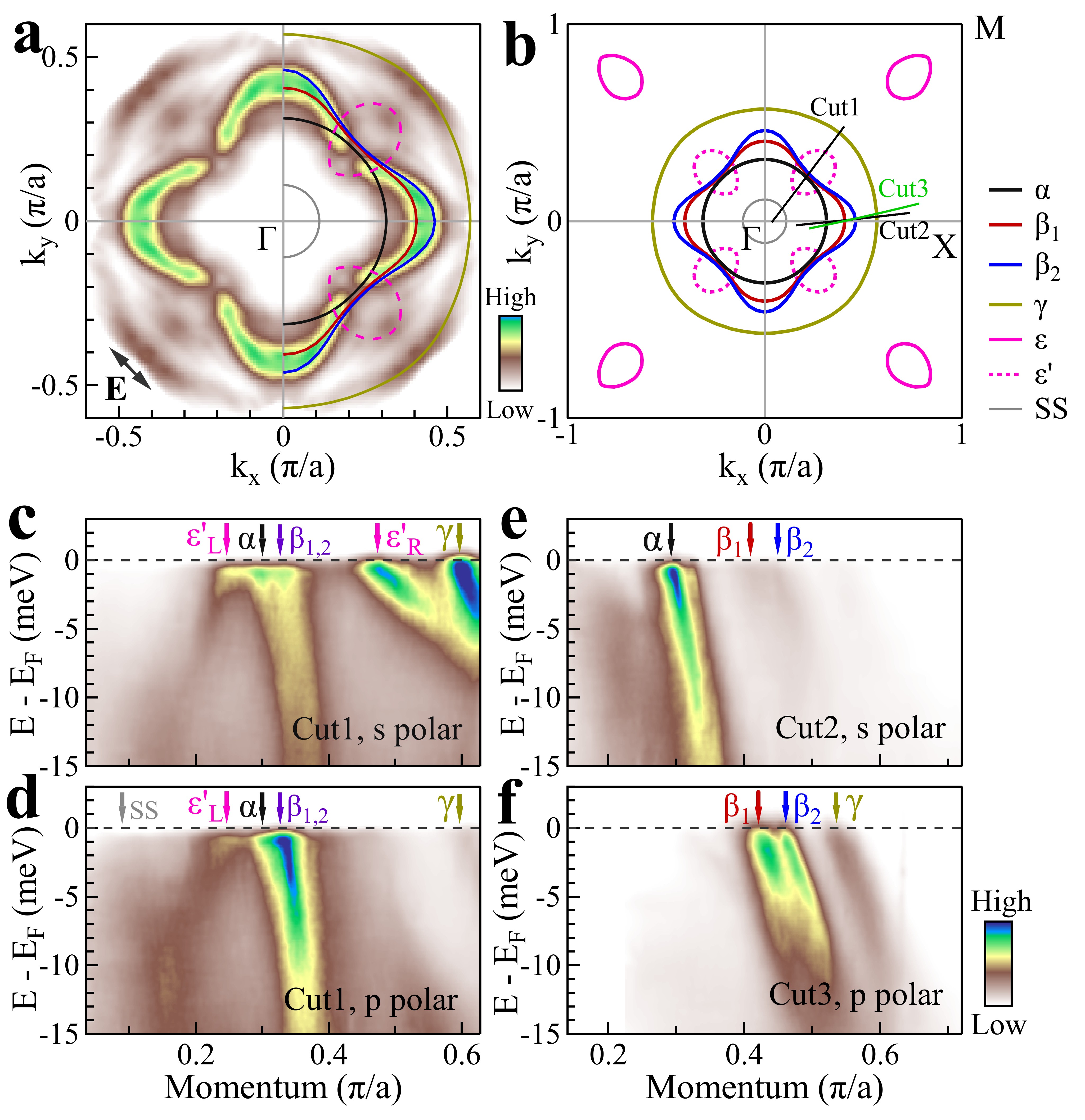}
\end{center}
\caption{{\bf Fermi surface and band structures of KFe$_2$As$_2$ measured at 0.9\,K.}  \textbf{a}, Fermi surface mapping near the zone center $\Gamma$ point. It is obtained by integrating the spectral intensity within $\pm$5\,meV with respect to the Fermi level and symmetrized assuming fourfold symmetry.
\textbf{b}, Measured Fermi surface obtained by analyzing the Fermi surface mappings and band structures measured under different polarization geometries (Fig. S2). The four $\varepsilon'$ pockets around $\Gamma$ are produced from folding the four $\varepsilon$ pockets around M due to $\sqrt2 \times \sqrt2$ surface reconstruction.
\textbf{c-f}, Typical band structures measured along different momentum cuts under different polarization geometries. The location of the momentum cuts is marked in \textbf{b}. The observed bands are marked by arrows with different colors.
}
\label{FS}
\end{figure*}

\vspace{3mm}
Figure \ref{FS} shows the measured Fermi surface and band structures of KFe$_2$As$_2$. As a typical multi-orbital system, we find that strong photoemission matrix element effects are involved in the ARPES measurements of KFe$_2$As$_2$. Therefore, in order to fully resolve the Fermi surface and band structures, we carried out ARPES measurements using different light polarizations (Fig. S2 in Supplementary Materials). Combining the Fermi surface mapping (Fig. \ref{FS}a) and the associated band structures (Fig. \ref{FS}c-f) measured under various polarization geometries, we have obtained the complete Fermi surface picture of KFe$_2$As$_2$, as shown in Fig. \ref{FS}b. It consists of three main Fermi surface sheets around $\Gamma$ ($\alpha$, $\beta$ and $\gamma$ in Fig. \ref{FS}b), a small surface state sheet around $\Gamma$ (SS in Fig. \ref{FS}b) and four hole pockets around M ($\varepsilon$ in Fig. \ref{FS}b). The Fermi surface and band structures of KFe$_2$As$_2$ we obtained from our laser ARPES measurements are consistent with the previous results\cite{HDing2009TSato, HHisatomo2014TYoshida}. They are much more clearly resolved than those in the previous laser ARPES measurement\cite{SShin2012KOkazaki}. With the better instrumental resolutions, we find that the $\beta$ Fermi surface further splits into two sheets ($\beta_1$ and $\beta_2$ in Fig. \ref{FS}b) with the maximum splitting along $\Gamma$-X, as seen from the band structure in Fig. \ref{FS}f. Such a splitting is not expected from the band structure calculations\cite{ZPYin2011}; it was not observed in the previous ARPES measurements either\cite{HDing2009TSato, SShin2012KOkazaki, HHisatomo2014TYoshida, HHWen2015DLFang}. Whether it comes from different k$_z$s or from surface effect needs further investigations.

Although laser ARPES is advantageous in taking high quality data with superhigh resolution, it has a limitation in the covered momentum space due to its lower photon energy\cite{XJZhou2018, SShin2012KOkazaki}. Particularly for the iron-based superconductors, laser ARPES can cover mainly the zone center region but cannot reach the zone corner M points\cite{XJZhou2018, SShin2012KOkazaki}. It is interesting and significant that in our present study we can observe the Fermi surface and band structures around M that are folded to the region around $\Gamma$. Under a proper polarization geometry, four Fermi pockets can be observed around $\Gamma$ which are marked by dashed pink lines and labeled as $\varepsilon'$ in Fig. \ref{FS}a. The corresponding band structures ($\varepsilon'_L$ and $\varepsilon'_R$ in Fig. \ref{FS}c and \ref{FS}d) can be clearly resolved and their spectral intensity is comparable to that of $\alpha$, $\beta$ and $\gamma$ bands. These four $\varepsilon'$ pockets can be well attributed to the four $\varepsilon$ hole pockets around M that are folded to the $\Gamma$ point by the $\sqrt{2}\times\sqrt{2}$ surface reconstruction (Fig. S3 in Supplementary Materials)\cite{HHWen2015DLFang, XJZhou2021YQCai2}. In (Ba$_{1-x}$K$_x$)Fe$_2$As$_2$ system, due to surface reconstruction, the band folding between $\Gamma$ and M points is commonly observed \cite{SShin2017TShimojima, XJZhou2021YQCai2, XJZhou2021YQCai}. Such a band folding makes it possible to study the electronic structure and superconducting gap of KFe$_2$As$_2$ in the full momentum space by laser ARPES.

\begin{figure*}[tbp]
\begin{center}
\includegraphics[width=1\textwidth,angle=0]{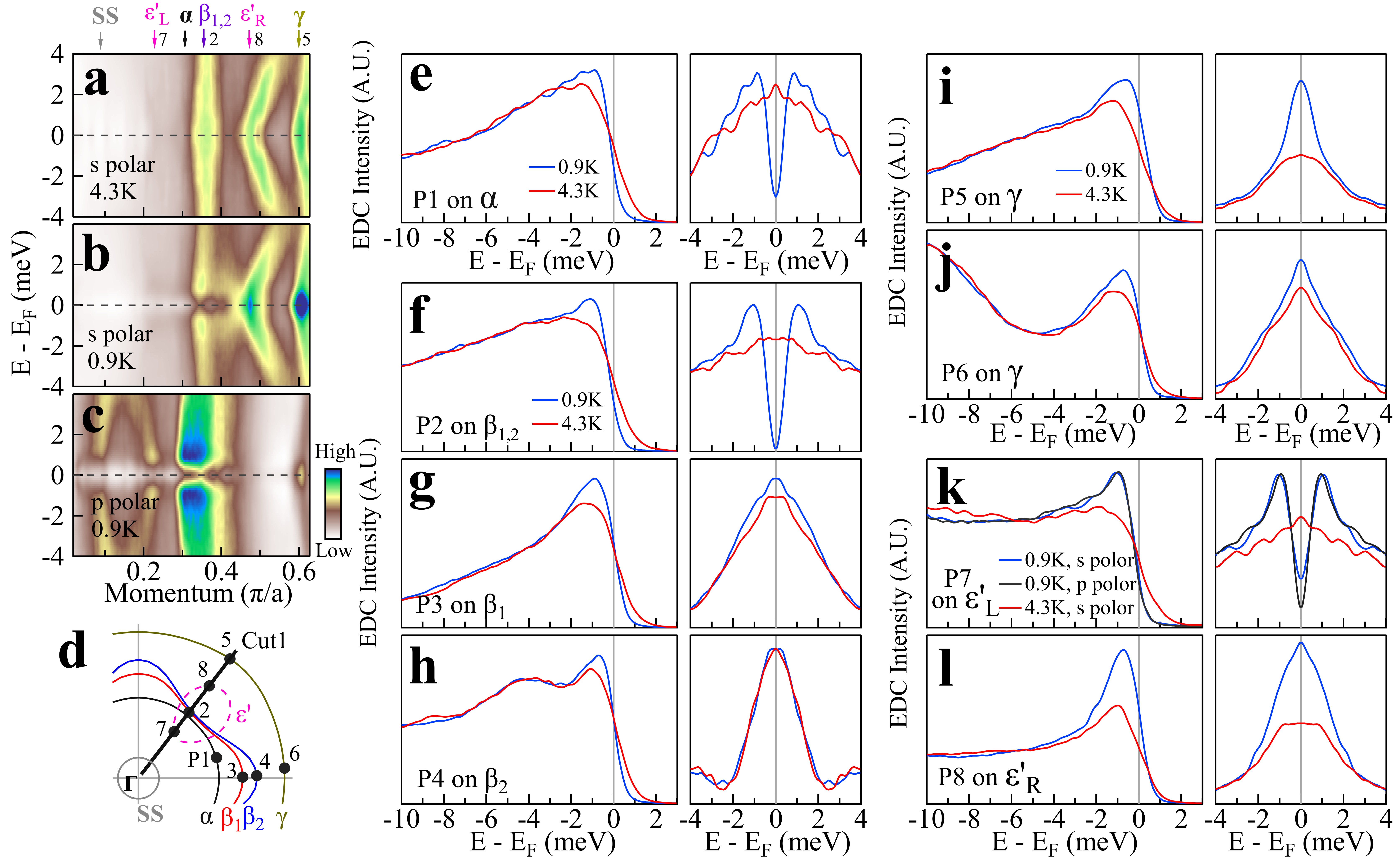}
\end{center}
\caption{{\bf Superconducting gap opening along different Fermi surface sheets by measuring photoemmission spectra in the normal and superconducting states.}
\textbf{a-c}, Band structures measured along the same momentum cut at different temperatures under different polarization geometries. The location of the momentum cut (Cut1) is marked by the black line in \textbf{d}. \textbf{a} is measured in the normal state at 4.3\,K under the s polarization. \textbf{b} is the same as \textbf{a} but measured at 0.9\,K in the superconducting state. \textbf{c} is the same as \textbf{b} but measured under the p polarization. To directly visualize the gap opening, these images are obtained by symmetrizing the original data with respect to the Fermi level. \textbf{d}, Fermi surface of KFe$_2$As$_2$ marked with the momentum cut used in \textbf{a-c} and the Fermi momentum points P1-P8 used in \textbf{e-l}. \textbf{e}, EDCs (left panel) and the corresponding symmetrized EDCs (right panel) measured at the Fermi momentum P1 along the $\alpha$ Fermi surface in the normal (4.3\,K) and superconducting (0.9\,K) states. The location of the P1 point is marked in \textbf{d}. \textbf{f-h}, Same as \textbf{e} but measured at the Fermi momenta P2 (\textbf{f}), P3 (\textbf{g}) and P4 (\textbf{h}) along the $\beta$ Fermi surface sheets ($\beta_1$ and $\beta_2$). \textbf{i-j}, Same as \textbf{e} but measured at the Fermi momenta P5 (\textbf{i}) and P6 (\textbf{j}) along the $\gamma$ Fermi surface sheet. \textbf{k-l}, Same as \textbf{e} but measured at the Fermi momenta P7 (\textbf{k}) and P8 (\textbf{l}) along the $\varepsilon'$ Fermi surface sheet. To facilitate a comparison, in \textbf{k}, the two EDCs measured at 0.9\,K under s and p polarizations are normalized in intensity near the Fermi level.
}
\label{EDCT}
\end{figure*}

Figure \ref{EDCT} highlights the band structures and photoemission spectra measured in the normal and superconducting states. Clear quasiparticle peaks are observed in the photoemission spectra (energy distribution curves, EDCs) measured along all the Fermi surface sheets at 4.3\,K in the normal state (red curves in left panels of Fig. \ref{EDCT}e-l). Superconducting coherence peaks develop in some EDCs in the superconducting state at 0.9\,K (blue curve in left panels of Fig. \ref{EDCT}e,f,k). The observed peaks are rather sharp with a width (full width at half maximum, FWHM) of 1$\sim$2 meV. This is in a strong contrast to the previous results where no EDC peaks are clearly observed\cite{SShin2012KOkazaki}. In order to visualize the gap opening, band symmetrization or EDC symmetrization with respect to the Fermi level is usually carried out to remove the Fermi distribution function\cite{JCCampuzano1998Norman}. In this case, the gap opening corresponds to the spectral weight suppression at the Fermi level. As visualized in Fig. \ref{EDCT}b,c, in the superconducting state at 0.9\,K, the spectral weight at the Fermi level for the $\alpha$, $\beta$, $\varepsilon'_L$ and SS bands is clearly suppressed indicating the opening of the superconducting gap. In the symmetrized EDCs measured at 0.9\,K (blue curves in the right panels of Fig. \ref{EDCT}e-l), the spectral weight suppression at the Fermi level is observed for P1 on $\alpha$ band (Fig. \ref{EDCT}e), P2 on $\beta$ band (Fig. \ref{EDCT}f) and P7 on $\varepsilon'_L$ band (Fig. \ref{EDCT}k), signaling the superconducting gap opening. The gap size can be determined from the energy difference between the two peaks in the symmetrized EDCs. On the other hand, the symmetrized EDCs in the superconducting state exhibit a peak at the Fermi level at other momentum points (blue curves in Fig. \ref{EDCT}g,h,i,j,l) indicating that no superconducting gap opening is detected within our experimental precision. In all the symmetrized EDCs measured at 4.3\,K in the normal state (red curves in right panels of Fig. \ref{EDCT}e-l), there is no signature of gap opening observed, excluding the formation of pseudogap in KFe$_2$As$_2$.

\begin{figure*}[tbp]
\begin{center}
\includegraphics[width=1\textwidth,angle=0]{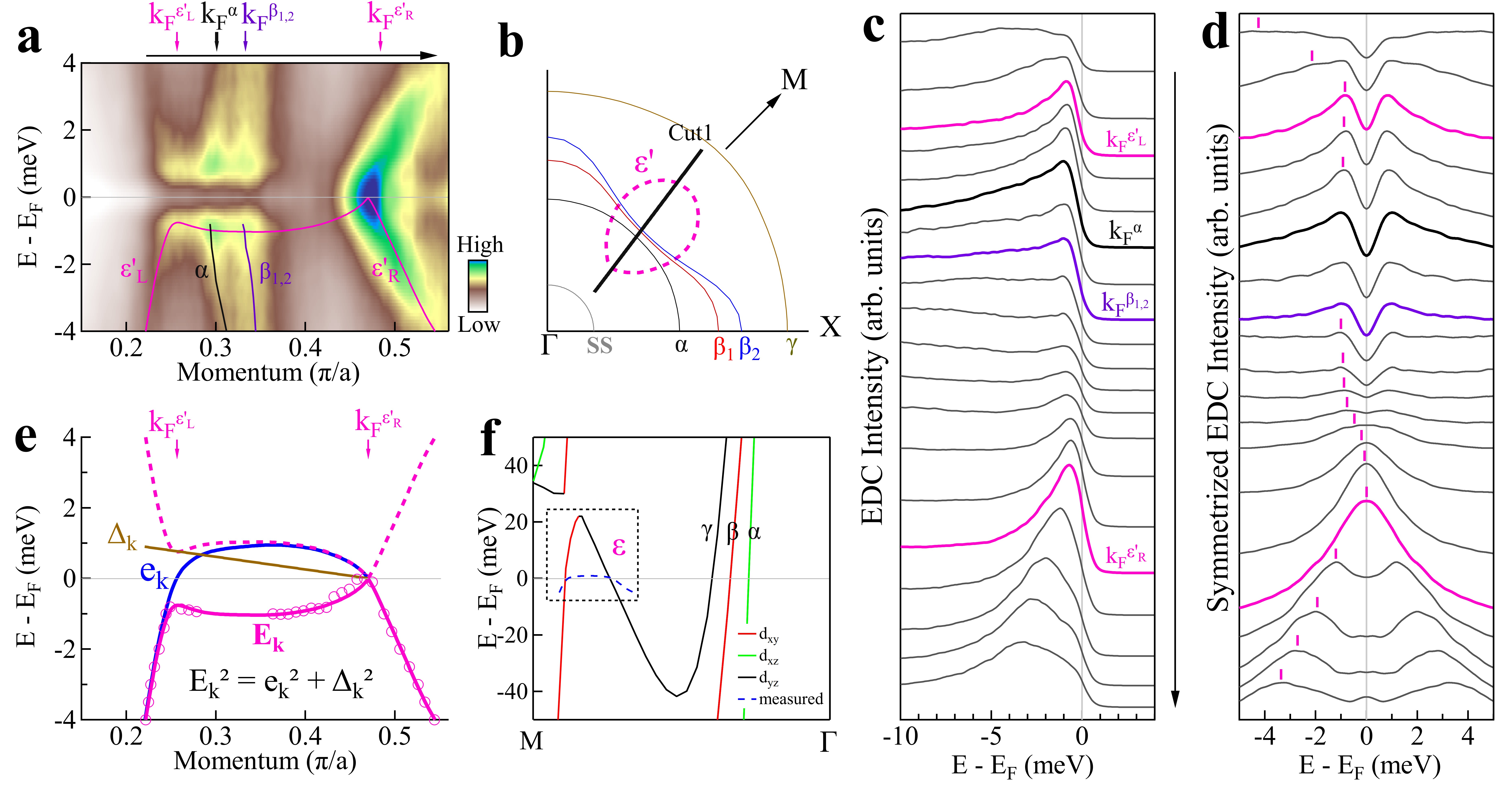}
\end{center}
\caption{{\bf Detailed analysis of the $\varepsilon$ band in the superconducting state.} \textbf{a}, Band structure measured along a momentum cut that is nearly parallel to the $\Gamma$-M direction. The location of the momentum cut (Cut1) is marked by the black line in \textbf{b}. To directly visualize the gap opening and remove the effect of the Fermi cutoff, the image is obtained by symmetrizing the original data with respect to the Fermi level. The observed bands, $\alpha$, $\beta_{1,2}$ and $\varepsilon'$, are marked by black, purple and pink lines, respectively. \textbf{b}, Fermi surface of KFe$_2$As$_2$ marked with the momentum cut (Cut1) used in \textbf{a}. \textbf{c}, Original EDCs from the original data in \textbf{a} before symmetrization in the momentum range marked by the black line with an arrow on top of \textbf{a}. The EDCs at the Fermi momenta of the $\alpha$, $\beta_{1,2}$ and $\varepsilon'$ bands are marked by colors. \textbf{d}, The corresponding symmetrized EDCs obtained from \textbf{c}. The EDC peaks corresponding to $\varepsilon'$ band and its Bogoliubov band are marked by pink ticks. \textbf{e}, Quantitative determination of the $\varepsilon'$ band dispersion (pink circles) in the superconducting state from \textbf{a} and \textbf{d}, which is described by continuous pink lines (E$_k$). Since the gap size is zero at the right Fermi momentum (k$_F^{\varepsilon'_R}$) and 0.8\,meV at the left Fermi momentum (k$_F^{\varepsilon'_L}$), a linear variation of the gap, $\Delta_k$, between the Fermi momenta is assumed (brown line). The dispersion in the normal state $\varepsilon_k$ (blue cueve) can then be derived using the equation E$_k^2$ = $e_k^2$ + $\Delta_k^2$ from BCS theory\cite{JBardeen1957}. \textbf{f}, The calculated band structure shows where the measured $\varepsilon$ band is located with a black dashed frame. The measured $\varepsilon$ band is also plotted (blue dashed line) for a direct comparison.
   }
\label{Epsilon2}
\end{figure*}

It was found that KFe$_2$As$_2$ exhibits heavy Fermion behaviours manifested by its large Sommerfeld coefficient and heavy effective mass\cite{UShinya2010TTaichi, XFWang2011JSKim, LBoeri2012MAbdel-Hafiez, SUji2013TTerashima, CMeingast2013FHardy, HVLohneysen2016FEilers}. Our present high resolution band structure measurements provide key information in understanding its origin. Fig. \ref{Epsilon2}a shows a detailed band structure of the $\varepsilon'$ Fermi surface measured along $\Gamma$-M direction at 0.9\,K. The corresponding EDCs and symmetrized EDCs are shown in Fig. \ref{Epsilon2}c and \ref{Epsilon2}d, respectively. The band structure of $\varepsilon'$ in the superconducting state, E$_k$, can be directly extracted from the symmetrized EDCs (Fig. \ref{Epsilon2}d) and plotted in Fig. \ref{Epsilon2}e as pink circles. The superconducting gap can also be determined directly from the symmetrized EDCs (Fig. \ref{Epsilon2}d) which are 0.8\,meV and 0 for the left and right Fermi momenta of the $\varepsilon'$ band (k$_F^{\varepsilon'_L}$ and k$_F^{\varepsilon'_R}$), respectively. Assuming a linear variation of the superconducting gap between the two Fermi momenta ($\Delta_k$), we obtain the normal state $\varepsilon'$ band structure (e$_k$) by using the BCS relation E$_k^2$ = e$_k^2$ + $\Delta_k^2$, as plotted by the blue line in Fig. \ref{Epsilon2}e. The normal state $\varepsilon'$ band is strikingly flat and its band top is only $\sim$1\,meV above the Fermi level. It is natural to expect that such a band will contribute significantly to the transport and thermodynamic properties. From the band structures in Fig. \ref{Epsilon2}e, we can also directly determine the effective mass of the $\varepsilon'$ Fermi surface. The effective mass obtained from the left side and the right side of the $\varepsilon'$ band is 3.2m$_e$ and 13.8m$_e$, respectively, with m$_e$ being the free electron mass. Such a dramatic anisotropy of the effective mass is understandable because the two sides of the $\varepsilon'$ band come from different orbitals as seen in Fig. \ref{Epsilon2}f. In particular, when compared with the band structure calculations (Fig. \ref{Epsilon2}f)\cite{UShinya2010TTaichi}, the effective mass for the left side and right side $\varepsilon$ band reaches 16m$_b$ and 8.1m$_b$ with m$_b$ being the corresponding band masses. The Fermi energy of the $\varepsilon$ band shrinks dramatically from the calculated $\sim$20\,meV to the measured $\sim$1\,meV.  Such a huge band renormalization of the $\varepsilon$ band is among the strongest that have been observed in the iron-based superconductors\cite{FBaumberger2010ATamai}.

Our detailed band structure measurements indicate that the large specific heat and the associated heavy Fermion behaviours discovered in KFe$_2$As$_2$\cite{UShinya2010TTaichi, XFWang2011JSKim, LBoeri2012MAbdel-Hafiez, SUji2013TTerashima, CMeingast2013FHardy, HVLohneysen2016FEilers, XHChen2016YPWu} originate from the strong band renormalizations, particularly the $\varepsilon$ band. Similar to the $\varepsilon'$ band in Fig. \ref{Epsilon2}e,f, we carried out band structure analysis of the $\alpha$, $\beta$ and $\gamma$ bands along different directions (Fig. S4 in Supplementary Materials) to extract the effective mass and the Sommerfeld coefficient $\gamma$, as shown in Fig. S4e,f and listed in Table S1 in Supplementary Materials. All the bands exhibit strongly enhanced effective mass with the $\gamma$ band reaching over 20 times of m$_e$ (Fig. S4e and Table S1 in Supplementary Materials). When compared with the calculated band structures, the $\varepsilon$ band shows the strongest renormalization with m$^*$/m$_b$ over 15 (Fig. S4f and Table S1 in Supplementary Materials). The anomalously large specific heat found in KFe$_2$As$_2$ comes from the combined contributions of all these renormalized bands. In spite of its small Fermi surface area, the $\varepsilon$ band plays a significant role in contributing to the specific heat; nearly half of the Sommerfeld coefficient is contributed by the $\varepsilon$ band (Table S1 in Supplementary Materials).

\begin{figure*}[tbp]
\begin{center}
\includegraphics[width=1\textwidth,angle=0]{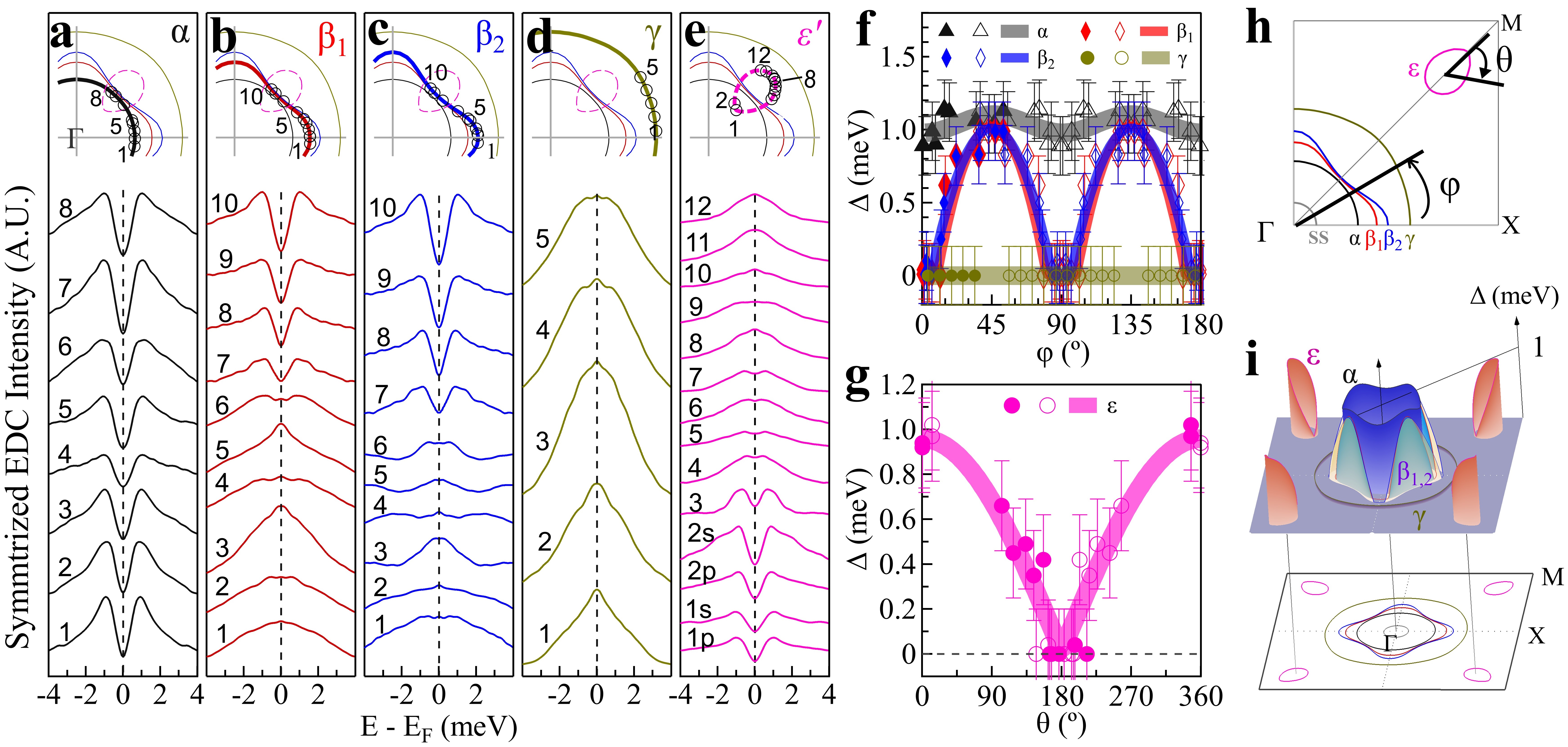}
\end{center}
\caption{{\bf Momentum-dependent superconducting gaps along all the Fermi surface sheets of KFe$_2$As$_2$.} \textbf{a-e}, Symmetrized EDCs along the $\alpha$ (\textbf{a}), $\beta_1$ (\textbf{b}), $\beta_2$ (\textbf{c}), $\gamma$ (\textbf{d}) and $\varepsilon'$ (\textbf{e}) Fermi surface sheets, respectively. The Fermi momentum positions are marked by empty circles in the inset Fermi surface. The details of extracting these symmetrized EDCs are described in Fig. S9-S13 in Supplementary Materials. \textbf{f-g}, Momentum-dependent superconducting gap as a function of the Fermi surface angle $\varphi$ for the $\alpha$, $\beta_1$, $\beta_2$ and $\gamma$ Fermi surface in \textbf{f} and $\theta$ for the $\varepsilon$ Fermi surface in \textbf{g}. The Fermi surface angles $\varphi$ and $\theta$ are defined in \textbf{h}. The solid symbols are derived from the symmetrized EDCs shown in \textbf{a-e} while the open symbols are obtained by symmetrization taking into account the fourfold symmetry. The thick lines are the fitted curves of the measured gaps. \textbf{i}, Three-dimensional plot of the superconducting gaps in KFe$_2$As$_2$. The corresponding Fermi surface is shown at the bottom.
   }
\label{Gapsym}
\end{figure*}

Now we come to the determination of the superconducting gap in KFe$_2$As$_2$. Since KFe$_2$As$_2$ is a typical multiorbital system and strong matrix element effects are involved in the ARPES measurements (Fig. \ref{FS}), in order to extract the superconducting gap along all the Fermi surface sheets, we carried out ARPES measurements along different momentum cuts under different polarization geometries and repeated the measurements on many KFe$_2$As$_2$ samples (Fig. S9-S13 in Supplementary Materials for the $\alpha$, $\beta$, $\gamma$, $\varepsilon'$ and SS, respectively). Fig. \ref{Gapsym}a-e shows the symmetrized EDCs along the $\alpha$, $\beta_1$, $\beta_2$, $\gamma$ and $\varepsilon'$ Fermi surface sheets. The gap size is determined by the peak position in the symmetrized EDCs\cite{JCCampuzano1998Norman} (for detailed discussion of the gap determination, see Fig. S5, S6 and S7 in Supplementary Materials). The measured superconducting gap along the $\alpha$, $\beta_1$, $\beta_2$ and $\gamma$ Fermi surface is presented in Fig. \ref{Gapsym}f and the gap along the $\varepsilon$ Fermi surface is shown in Fig. \ref{Gapsym}g. Fig. \ref{Gapsym}i shows a three-dimensional plot of the measured superconducting gap structure in KFe$_2$As$_2$. As seen in Fig. \ref{Gapsym}f, the $\alpha$ Fermi surface shows a large superconducting gap ($\sim$1.0\,meV) which is nearly isotropic with a small variation of $\pm$0.1\,meV. The superconducting gap of the $\beta$ Fermi surface ($\beta_1$ and $\beta_2$) is highly anisotropic. It is nearly zero along the $\Gamma$-X direction and reaches the maximum ($\sim$1.0\,meV) along the $\Gamma$-M direction. Along the entire $\gamma$ Fermi surface, the gap is nearly zero within our detection limit. As seen in Fig. \ref{Gapsym}g, the superconducting gap on the small $\varepsilon$ pocket is highly anisotropic. Along the $\Gamma$-M direction, it is the maximum near the tip close to M while it approaches zero on the other side away from M. The superconducting gap structure in KFe$_2$As$_2$ is quite distinct along different Fermi surface sheets.

We note that our results of the superconducting gap along the $\alpha$, $\beta$ and $\gamma$ Fermi surface around $\Gamma$ (Fig. \ref{Gapsym}f) show significant difference from those reported before\cite{SShin2012KOkazaki} (the reason of the discrepancy is discussed in Supplementary Materials). The superconducting gap along the $\varepsilon$ Fermi surface was not measured in the previous ARPES measurement\cite{SShin2012KOkazaki} and, for the first time, it has been measured here. Based on the previous ARPES result\cite{SShin2012KOkazaki}, it was proposed that the superconducting gap of KFe$_2$As$_2$ reverses its sign between the $\alpha$ and $\beta$ Fermi surface sheets around $\Gamma$\cite{AVChubukov2012SMaiti, AVChubukov2017OVafek}. Neutron scattering on KFe$_2$As$_2$ has found spin resonance in the superconducting state\cite{JZhao2020SDShen}. Since the spin resonance mode is a signature of the gap sign change, the observation was considered to be consistent with the s-wave pairing with the reversed sign between the two hole pockets around $\Gamma$\cite{JZhao2020SDShen}.

\begin{figure*}[tbp]
\begin{center}
\includegraphics[width=1\textwidth,angle=0]{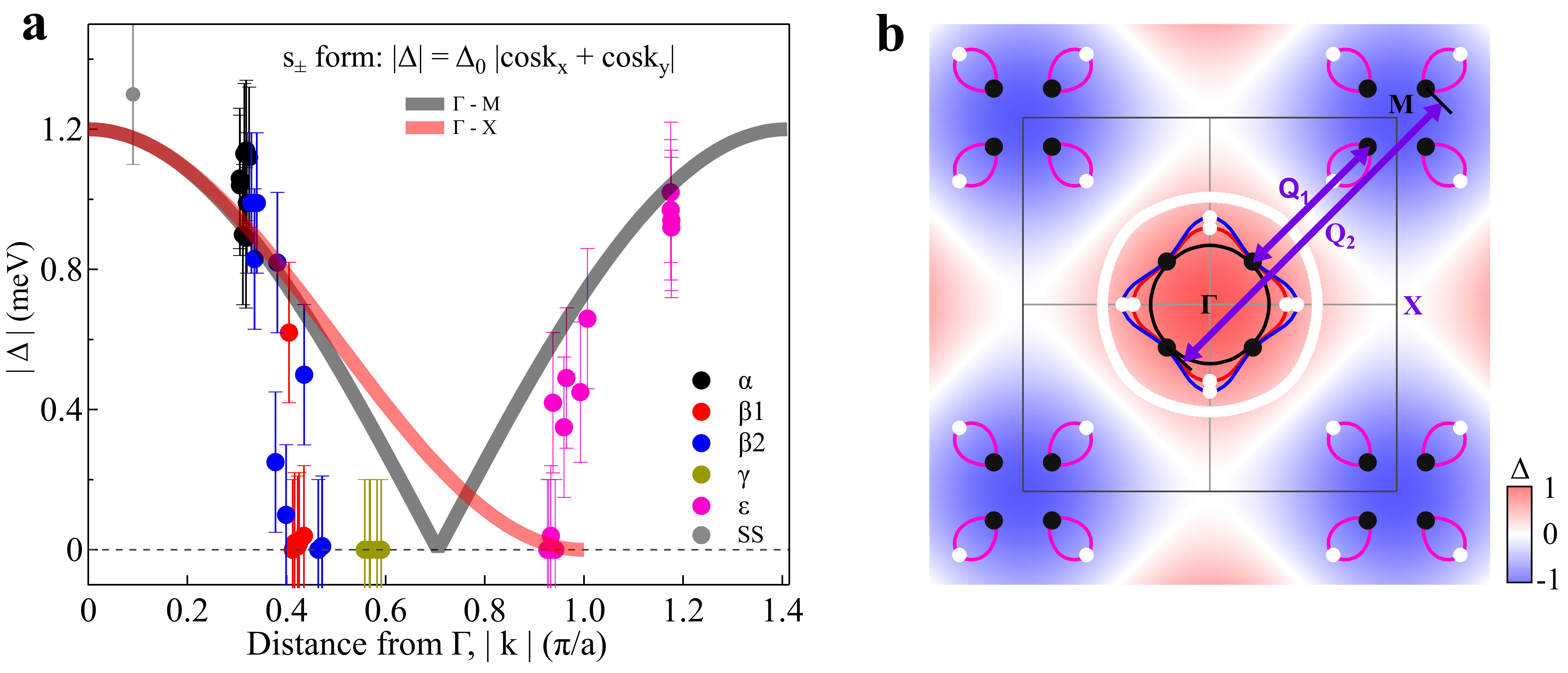}
\end{center}
\caption{{\bf Gap symmetry of KFe$_2$As$_2$.} \textbf{a}, The superconducting gap as a function of the distance between the Fermi momentum and the $\Gamma$ point on all the Fermi surface sheets. The expected gap size variation from the $\textsl{s}_\pm$ symmetry with the gap form $|$cosk$_x$+cosk$_y|$\cite{JPHu2012HDing} is plotted as a thick black line along $\Gamma$-M direction and as a thick red line along $\Gamma$-X direction.
\textbf{b}, Gap nodes and local gap maxima on the Fermi surface of KFe$_2$As$_2$. The Fermi surface are plotted on top of a two-dimensional image showing the gap function cosk$_x$+cosk$_y$ in the $\textsl{s}_\pm$ pairing symmetry where the positive gap sign, negative gap sign and the gap nodes are represented by red, blue and white colors, respectively. The measured gap nodes are marked by white circles while the gap maxima are marked by black circles. Two interband scattering wave vectors (\textbf{Q}$_1$ and \textbf{Q}$_2$) are marked by double-arrow purple lines. \textbf{Q}$_1$ and \textbf{Q}$_2$ connect the local gap maxima between the $\varepsilon$ band and ($\alpha$, $\beta$) bands.
   }                    
\label{gaps}
\end{figure*}

Our precise determination of the electronic structure and superconducting gap structure, particularly for the $\varepsilon$ Fermi surface around M, provides a new picture of the pairing symmetry in KFe$_2$As$_2$. We find that our results are more consistent with the $\textsl{s}_\pm$ pairing symmetry with the sign reversal between the $\Gamma$ and M points based on the following observations. First, the superconducting gap size along all the Fermi surface sheets is consistent with the simple $\textsl{s}_\pm$ form: {$|\Delta|$ = $\Delta_0 |$cosk$_x$+cosk$_y|$}, as shown in Fig. \ref{gaps}a. Second, precise nesting vectors along $\Gamma$-M can be determined from our measured Fermi surface (Fig. \ref{gaps}b). They are consistent with the previous ARPES measurements\cite{HHisatomo2014TYoshida} and show rather weak k$_z$ dependence\cite{HHisatomo2014TYoshida}. If the nestings connect the $\alpha$ and $\beta$ Fermi surface around $\Gamma$ as proposed before\cite{AVChubukov2012SMaiti, AVChubukov2017OVafek, JZhao2020SDShen}, the corresponding nesting vectors obtained from Fig. \ref{gaps}b are 0.61\,$\pi$/a and 2.21\,$\pi$/a along the $\Gamma$-M direction which significantly deviate from the two spin resonance wavevectors of 0.85\,$\pi$/a and 1.98\,$\pi$/a from neutron scattering measurements\cite{JZhao2020SDShen}. On the other hand, when the two vectors connect the best nested portion of the $\varepsilon$ Fermi surface around M and the $\alpha/\beta$ Fermi surface around $\Gamma$, as we identified in Fig. \ref{gaps}b, the obtained wavevectors of \textbf{Q}$_1$=0.87\,$\pi$/a and \textbf{Q}$_2$=1.96\,$\pi$/a have a perfect match to the two spin resonance wavevectors\cite{JZhao2020SDShen}. This indicates that the two regions connected by the nesting vectors have the reversed sign of the superconducting gap. This is consistent with the $\textsl{s}_\pm$ form {$\Delta$ = $\Delta_0$(cosk$_x$+cosk$_y$)}, as depicted in Fig. \ref{gaps}b. Third, the nesting vectors connect portions of the two Fermi surface that have the maximum superconducting gap (black solid circles in Fig. \ref{gaps}b). It has been found that the energy of the resonance mode E$_R$ and the superconducting gap $\Delta_{tot}$ follow a universal relation of E$_R$/$\Delta_{tot}$$\approx$0.64 in the iron-based superconductors with $\Delta_{tot}$ being the total superconducting gap summed on the two Fermi surfaces linked by the nesting vectors\cite{MGreven2009GYu}. The energy of the resonance mode in KFe$_2$As$_2$ estimated from $\Delta_{tot}$ = 1.0+1.0\,meV is 1.28\,meV which is in a good agreement with that measured in neutron scattering (1.2\,meV)\cite{JZhao2020SDShen}. Fourth, from the measured superconducting gap structure, we have identified the nearly zero gap points on the $\beta$, $\gamma$ and $\varepsilon$ Fermi surface, as marked by the white circles in Fig. \ref{gaps}b. In the $\textsl{s}_\pm$ gap symmetry, there are zero gap nodal regions along (0,$\pm\pi$)-($\pm\pi$,0) lines (Fig. \ref{gaps}b). It is interesting to note that the measured momentum region with nearly zero gap on the $\beta$ and $\varepsilon$ Fermi surface is located closest to the zero gap lines of the $\textsl{s}_\pm$ form. In particular, the $\gamma$ Fermi surface lies very close to the $\textsl{s}_\pm$ nodal lines and the superconducting gap is nearly zero along the entire Fermi surface.  These results indicate that the sign reversal $\textsl{s}_\pm$ pairing symmetry can be realized in the system with only hole pockets.  Here the sign change occurs between the hole pockets around $\Gamma$ and those around M. Previously the $\textsl{s}_\pm$ pairing symmetry was proposed and tested in the superconductors with hole pockets around $\Gamma$ and electron pockets around M\cite{IIMazin2008, HAoki2008KKuroki, JPHu2008KSeo, DHLee2009FWang, HTakagi2010THanaguri}. Neutron scatterings have observed spin fluctuations in KFe$_2$As$_2$\cite{KYamada2011CHLee, CHLee2016KHorigane, JZhao2020SDShen} and spin fluctuations exchange always leads to a repulsive interaction and can only realize sign-changing superconducting states\cite{IIMazin2011PJHirschfeld}. Therefore, our identification of the nodal $\textsl{s}_\pm$ pairing symmetry in KFe$_2$As$_2$ indicates that the spin fluctuations may play a dominant role in generating superconductivity even in a system with only hole pockets.

In summary, by taking super-high resolution laser ARPES measurements at ultra-low temperature, we have determined the complete electronic structure and superconducting gap structure of KFe$_2$As$_2$. We have uncovered the electronic origin of the large specific heat and heavy Fermion behaviours found in the system. Our observations point to a nodal s$_\pm$ pairing symmetry with a sign reversal between the zone center and the zone corners realized in the iron-based superconductors with only hole pockets. These results provide key information in understanding the pairing symmetry and the superconductivity mechanism in iron-based superconductors.

\newpage
\noindent{\bf Methods}\vspace{3mm}\\
\hspace*{6mm} High-quality single crystals of KFe$_2$As$_2$ were grown by the KAs flux method\cite{HEisaki2016KKihou}. The samples were characterized by magnetic susceptibility measurements (Fig. S1a in Supplementary Materials) and the measured T$_C$ is 3.7\,K with a narrow transition width of  $\sim$0.2\,K. The residual resistivity rate (R$_{300K}$/R$_{0K}$) of near 1000 is obtained from the resistivity measurement (Fig. S1b in Supplementary Materials). Both measurements indicate that our KFe$_2$As$_2$ single crystal samples have very high quality.

High-resolution angle-resolved photoemission measurements were carried out by using our laboratory-based ARPES system. It is equipped with a vacuum-ultraviolet (VUV) laser with a photon energy of h$\nu$ = 6.994\,eV and a hemispherical electron energy analyzer R8000 (Scienta-Omicron). The light polarization can be varied to get linear polarization along different directions and circular polarization. This system uses the $^3$He pumping technology which can cool the sample down to the lowest temperature of 0.8\,K. The energy resolution was set at 0.5\,meV and the angular resolution was $\sim$0.3$^\circ$ corresponding to 0.004\, ${\AA}$$^{-1}$ momentum resolution at the photon energy of 6.994\,eV. All the samples were cleaved \textit{in situ} at low temperature and measured in ultrahigh vacuum with a base pressure better than 1$\times$10$^{-10}$\,mbar. The Fermi level is carefully referenced by measuring polycrystalline gold which is well connected with the sample.


\vspace{3mm}
\noindent$^{*}$Corresponding author:  LZhao@iphy.ac.cn,  XJZhou@iphy.ac.cn.


\vspace{3mm}
\begin{thebibliography}{10}

\bibitem{HHosono2008YKamihara}
Y.~Kamihara, T.~Watanabe, M.~Hirano, and H.~Hosono.
\newblock Iron-based layered superconductor {La[O$_{1-x}$F$_x$]FeAs} (x =
  0.05-0.12) with {T$_c$} = 26 {K}.
\newblock {\em J. Am. Chem. Soc.}, 130(11):3296--3297, 2008.

\bibitem{ZXZhao2008ZARen}
Z.~A. Ren, W.~Lu, J.~Yang, W.~Yi, X.~L. Shen, C.~Zheng, G.~C. Che, X.~L. Dong,
  L.~L. Sun, F.~Zhou, and Z.~X. Zhao.
\newblock Superconductivity at 55 {K} in iron-based {F}-doped layered
  quaternary compound {Sm[O$_{1-x}$F$_x$]FeAs}.
\newblock {\em Chin. Phys. Lett.}, 25(6):2215, 2008.

\bibitem{DJohrendt2008MRotter}
M.~Rotter, M.~Pangerl, M.~Tegel, and D.~Johrendt.
\newblock Superconductivity and crystal structures of
  {(Ba$_{1-x}$K$_x$)Fe$_2$As$_2$ (x=0-1)}.
\newblock {\em Angew Chem Int Ed Engl}, 47(41):7949--52, 2008.

\bibitem{CQJin2008XCWang}
X.~C. Wang, Q.~Q. Liu, Y.~X. Lv, W.~B. Gao, L.~X. Yang, R.~C. Yu, F.~Y. Li, and
  C.~Q. Jin.
\newblock The superconductivity at 18 {K} in {LiFeAs} system.
\newblock {\em Solid State Commun.}, 148(11-12):538--540, 2008.

\bibitem{MWu2008FHsu}
F.~C. Hsu, J.~Y. Luo, K.~W. Yeh, T.~K. Chen, T.~W. Huang, P.~M. Wu, Y.~C. Lee,
  Y.~L. Huang, Y.~Y. Chu, D.~Ch. Yan, and M.~K. Wu.
\newblock Superconductivity in the {PbO}-type structure.
\newblock {\em Proc. Natl. Acad. Sci. U.S.A.}, 105(38):14262--14264, 2008.

\bibitem{IIMazin2011PJHirschfeld}
P.~J. Hirschfeld, M.~M. Korshunov, and I.~I. Mazin.
\newblock Gap symmetry and structure of {Fe}-based superconductors.
\newblock {\em Rep. Prog. Phys.}, 74(12):124508, 2011.

\bibitem{DHLee2011FWang}
F.~Wang and D.~H. Lee.
\newblock The electron-pairing mechanism of iron-based superconductors.
\newblock {\em Science}, 332(6026):200--204, 2011.

\bibitem{HHosono2015}
H.~Hosono and K.~Kuroki.
\newblock Iron-based superconductors: Current status of materials and pairing
  mechanism.
\newblock {\em Physica C}, 514:399--422, 2015.

\bibitem{RFernandes2022}
R.~M. Fernandes, A.~I. Coldea, H.~Ding, I.~R. Fisher, P.~J. Hirschfeld, and
  G.~Kotliar.
\newblock Iron pnictides and chalcogenides: a new paradigm for
  superconductivity.
\newblock {\em Nature}, 601(7891):35--44, 2022.

\bibitem{IIMazin2008}
I.~I. Mazin, D.~J. Singh, M.~D. Johannes, and M.~H. Du.
\newblock Unconventional superconductivity with a sign reversal in the order
  parameter of {LaFeAsO$_{1-x}$F$_x$}.
\newblock {\em Phys. Rev. Lett.}, 101(5):057003, 2008.

\bibitem{HAoki2008KKuroki}
K.~Kuroki, S.~Onari, R.~Arita, H.~Usui, Y.~Tanaka, H.~Kontani, and H.~Aoki.
\newblock Unconventional pairing originating from the disconnected {Fermi}
  surfaces of superconducting {LaFeAsO$_{1-x}$F$_x$}.
\newblock {\em Phys. Rev. Lett.}, 101(8):087004, 2008.

\bibitem{AElihu2008SQimiao}
Q.~Si and E.~Abrahams.
\newblock Strong correlations and magnetic frustration in the high {T$_c$} iron
  pnictides.
\newblock {\em Phys. Rev. Lett.}, 101(7):076401, 2008.

\bibitem{JPHu2008KSeo}
K.~Seo, B.~A. Bernevig, and J.~Hu.
\newblock Pairing symmetry in a two-orbital exchange coupling model of
  oxypnictides.
\newblock {\em Phys. Rev. Lett.}, 101(20):206404, 2008.

\bibitem{SOnari2010HKontani}
H.~Kontani and S.~Onari.
\newblock Orbital-fluctuation-mediated superconductivity in iron pnictides:
  analysis of the five-orbital {Hubbard-Holstein} model.
\newblock {\em Phys. Rev. Lett.}, 104(15):157001, 2010.

\bibitem{YOno2010YYanagi2}
Y.~Yanagi, Y.~Yamakawa, and Y.~Ono.
\newblock Two types of s-wave pairing due to magnetic and orbital fluctuations
  in the two-dimensional 16-band d-p model for iron-based superconductors.
\newblock {\em Phys. Rev. B}, 81(5):054518, 2010.

\bibitem{YOno2010YYanagi}
Y.~Yanagi, Y.~Yamakawa, N.~Adachi, and Y.~Ono.
\newblock Cooperative effects of {Coulomb} and electron-phonon interactions in
  the two-dimensional 16-band d-p model for iron-based superconductors.
\newblock {\em Phys. Rev. B}, 82(6):064518, 2010.

\bibitem{DLFeng2011YZhang}
Y.~Zhang, L.~X. Yang, M.~Xu, Z.~R. Ye, F.~Chen, C.~He, H.~C. Xu, J.~Jiang,
  B.~P. Xie, J.~J. Ying, X.~F. Wang, X.~H. Chen, J.~P. Hu, M.~Matsunami,
  S.~Kimura, and D.~L. Feng.
\newblock Nodeless superconducting gap in {A$_x$Fe$_2$Se$_2$ (A=K,Cs)} revealed
  by angle-resolved photoemission spectroscopy.
\newblock {\em Nat. Mater.}, 10(4):273--277, 2011.

\bibitem{HDing2011TQian}
T.~Qian, X.~P. Wang, W.~C. Jin, P.~Zhang, P.~Richard, G.~Xu, X.~Dai, Z.~Fang,
  J.~G. Guo, X.~L. Chen, and H.~Ding.
\newblock Absence of a holelike {Fermi} surface for the iron-based
  {K$_{0.8}$F$_{1.7}$Se$_2$} superconductor revealed by angle-resolved
  photoemission spectroscopy.
\newblock {\em Phys. Rev. Lett.}, 106(18):187001, 2011.

\bibitem{XJZhou2011DXMou}
D.~Mou, S.~Liu, X.~Jia, J.~He, Y.~Peng, L.~Zhao, L.~Yu, G.~Liu, S.~He, X.~Dong,
  J.~Zhang, H.~Wang, C.~Dong, M.~Fang, X.~Wang, Q.~Peng, Z.~Wang, S.~Zhang,
  F.~Yang, Z.~Xu, C.~Chen, and X.~J. Zhou.
\newblock Distinct {Fermi surface} topology and nodeless superconducting gap in
  a {(Tl$_{0.58}$Rb$_{0.42}$)Fe$_{1.72}$Se$_2$} superconductor.
\newblock {\em Phys. Rev. Lett.}, 106(10):107001, 2011.

\bibitem{XJZhou2012DFLiu}
D.~Liu, W.~Zhang, D.~Mou, J.~He, Y.~B. Ou, Q.~Y. Wang, Z.~Li, L.~Wang, L.~Zhao,
  S.~He, Y.~Peng, X.~Liu, C.~Chen, L.~Yu, G.~Liu, X.~Dong, J.~Zhang, C.~Chen,
  Z.~Xu, J.~Hu, X.~Chen, X.~Ma, Q.~Xue, and X.~J. Zhou.
\newblock {Electronic origin of high-temperature superconductivity in
  single-layer FeSe superconductor}.
\newblock {\em Nat. Commun.}, 3:931, 2012.

\bibitem{XJZhou2016LZhao}
L.~Zhao, A.~Liang, D.~Yuan, Y.~Hu, D.~Liu, J.~Huang, S.~He, B.~Shen, Y.~Xu,
  X.~Liu, L.~Yu, G.~Liu, H.~Zhou, Y.~Huang, X.~Dong, F.~Zhou, K.~Liu, Z.~Lu,
  Z.~Zhao, C.~Chen, Z.~Xu, and X.~J. Zhou.
\newblock {Common electronic origin of superconductivity in (Li,Fe)OHFeSe bulk
  superconductor and single-layer FeSe/SrTiO$_3$ films}.
\newblock {\em Nat. Commun.}, 7:10608, 2016.

\bibitem{DJScalapino2011TAMaier}
T.~A. Maier, S.~Graser, P.~J. Hirschfeld, and D.~J. Scalapino.
\newblock d-wave pairing from spin fluctuations in the {K$_x$Fe$_{2-y}$Se$_2$}
  superconductors.
\newblock {\em Phys. Rev. B}, 83(10):100515, 2011.

\bibitem{DHLee2011FWang2}
F.~Wang, F.~Yang, M.~Gao, Z.~Y. Lu, T.~Xiang, and D.~H. Lee.
\newblock The electron pairing of {K$_x$Fe$_{2-y}$Se$_2$}.
\newblock {\em EPL (Europhysics Letters)}, 93(5):57003, 2011.

\bibitem{AVChubukov2012MKhodas}
M.~Khodas and A.~V. Chubukov.
\newblock Interpocket pairing and gap symmetry in {Fe}-based superconductors
  with only electron pockets.
\newblock {\em Phys. Rev. Lett.}, 108(24):247003, 2012.

\bibitem{BABernevig2011RThomale}
R.~Thomale, C.~Platt, W.~Hanke, J.~Hu, and B.~A. Bernevig.
\newblock Exotic d-wave superconducting state of strongly hole-doped
  {K$_x$Ba$_{1-x}$Fe$_2$As$_2$}.
\newblock {\em Phys. Rev. Lett.}, 107(11):117001, 2011.

\bibitem{AVChubukov2011SMaiti2}
S.~Maiti, M.~M. Korshunov, T.~A. Maier, P.~J. Hirschfeld, and A.~V. Chubukov.
\newblock Evolution of the superconducting state of {Fe}-based compounds with
  doping.
\newblock {\em Phys. Rev. Lett.}, 107(14):147002, 2011.

\bibitem{KKuroki2011KSuzuki}
K.~Suzuki, H.~Usui, and K.~Kuroki.
\newblock Spin fluctuations and unconventional pairing in {KFe$_2$As$_2$}.
\newblock {\em Phys. Rev. B}, 84(14):144514, 2011.

\bibitem{AVChubukov2012SMaiti}
S.~Maiti, M.~M. Korshunov, and A.~V. Chubukov.
\newblock Gap symmetry in {KFe${}_{2}$As${}_{2}$} and the
  $\mathrm{cos}${4}$\ensuremath{\theta}$ gap component in {LiFeAs}.
\newblock {\em Phys. Rev. B}, 85(1):014511, 2012.

\bibitem{AVChubukov2017OVafek}
O.~Vafek and A.~V. Chubukov.
\newblock Hund interaction, spin-orbit coupling, and the mechanism of
  superconductivity in strongly hole-doped iron pnictides.
\newblock {\em Phys. Rev. Lett.}, 118(8):087003, 2017.

\bibitem{HEisaki2009HFukazawa}
H.~Fukazawa, Y.~Yamada, K.~Kondo, T.~Saito, Y.~Kohori, K.~Kuga, Y.~Matsumoto,
  S.~Nakatsuji, H.~Kito, P.~M.~Shirage, K.~Kihou, N.~Takeshita, C.~H. Lee,
  A.~Iyo, and H.~Eisaki.
\newblock Possible multiple gap superconductivity with line nodes in heavily
  hole-doped superconductor {KFe$_2$As$_2$} studied by $^{75}${As} nuclear
  quadrupole resonance and specific heat.
\newblock {\em J. Phys. Soc. Jpn.}, 78(8):083712, 2009.

\bibitem{SYLi2010JKDongPRL}
J.~K. Dong, S.~Y. Zhou, T.~Y. Guan, H.~Zhang, Y.~F. Dai, X.~Qiu, X.~F. Wang,
  Y.~He, X.~H. Chen, and S.~Y. Li.
\newblock Quantum criticality and nodal superconductivity in the {FeAs}-based
  superconductor {KFe$_{2}$As$_{2}$}.
\newblock {\em Phys. Rev. Lett.}, 104(8):087005, 2010.

\bibitem{TShibauchi2010KHashimoto}
K.~Hashimoto, A.~Serafin, S.~Tonegawa, R.~Katsumata, R.~Okazaki, T.~Saito,
  H.~Fukazawa, Y.~Kohori, K.~Kihou, C.~H. Lee, A.~Iyo, H.~Eisaki, H.~Ikeda,
  Y.~Matsuda, A.~Carrington, and T.~Shibauchi.
\newblock {Evidence for superconducting gap nodes in the zone-centered hole
  bands of {KFe$_{2}$As$_{2}$} from magnetic penetration-depth measurements}.
\newblock {\em Phys. Rev. B}, 82(1):014526, 2010.

\bibitem{LTaillefer2012JPhReid}
J.~P. Reid, M.~A. Tanatar, A.~Juneau-Fecteau, R.~T. Gordon, S.~R. de~Cotret,
  N.~Doiron-Leyraud, T.~Saito, H.~Fukazawa, Y.~Kohori, K.~Kihou, C.~H. Lee,
  A.~Iyo, H.~Eisaki, R.~Prozorov, and L.~Taillefer.
\newblock Universal heat conduction in the iron arsenide superconductor
  {KFe$_{2}$As$_{2}$}: evidence of a d-wave state.
\newblock {\em Phys. Rev. Lett.}, 109(8):087001, 2012.

\bibitem{SShin2012KOkazaki}
K.~Okazaki, Y.~Ota, Y.~Kotani, W.~Malaeb, Y.~Ishida, T.~Shimojima, T.~Kiss,
  S.~Watanabe, C.~T. Chen, K.~Kihou, C.~H. Lee, A.~Iyo, H.~Eisaki, T.~Saito,
  H.~Fukazawa, Y.~Kohori, K.~Hashimoto, T.~Shibauchi, Y.~Matsuda, H.~Ikeda,
  H.~Miyahara, R.~Arita, A.~Chainani, and S.~Shin.
\newblock Octet-line node structure of superconducting order parameter in
  {KFe$_{2}$As$_{2}$}.
\newblock {\em Science}, 337(6100):1314--1317, 2012.

\bibitem{HDing2013NXu_BKFA}
N.~Xu, P.~Richard, X.~Shi, A.~van Roekeghem, T.~Qian, E.~Razzoli, E.~Rienks,
  G.~F. Chen, E.~Ieki, K.~Nakayama, T.~Sato, T.~Takahashi, M.~Shi, and H.~Ding.
\newblock Possible nodal superconducting gap and {Lifshitz} transition in
  heavily hole-doped {Ba$_{0.1}$K$_{0.9}$Fe$_2$As$_2$}.
\newblock {\em Phys. Rev. B}, 88(22):220508, 2013.

\bibitem{SLDrechsler2013MAbdel-Hafiez}
M.~Abdel-Hafiez, V.~Grinenko, S.~Aswartham, I.~Morozov, M.~Roslova,
  O.~Vakaliuk, S.~Johnston, D.~V. Efremov, J.~van~den Brink, H.~Rosner,
  M.~Kumar, C.~Hess, S.~Wurmehl, A.~U.~B. Wolter, B.~Buchner, E.~L. Green,
  J.~Wosnitza, P.~Vogt, A.~Reifenberger, C.~Enss, M.~Hempel, R.~Klingeler, and
  S.~L. Drechsler.
\newblock Evidence of $d$-wave superconductivity in
  {K${}_{1\ensuremath{-}x}$Na${}_{x}$Fe${}_{2}$As${}_{2}$} ($x=0,0.1$) single
  crystals from low-temperature specific-heat measurements.
\newblock {\em Phys. Rev. B}, 87(18):180507, 2013.

\bibitem{YMatsuda2014DWatanabe}
D.~Watanabe, T.~Yamashita, Y.~Kawamoto, S.~Kurata, Y.~Mizukami, T.~Ohta,
  S.~Kasahara, M.~Yamashita, T.~Saito, H.~Fukazawa, Y.~Kohori, S.~Ishida,
  K.~Kihou, C.~H. Lee, A.~Iyo, H.~Eisaki, A.~B. Vorontsov, T.~Shibauchi, and
  Y.~Matsuda.
\newblock {Doping evolution of the quasiparticle excitations in heavily
  hole-doped Ba${}_{1\ensuremath{-}x}$K${}_{x}$Fe${}_{2}$As${}_{2}$: A possible
  superconducting gap with sign-reversal between hole pockets}.
\newblock {\em Phys. Rev. B}, 89(11):115112, 2014.

\bibitem{RProzorov2016KCho}
C.~Kyuil, K.~Marcin, T.~Serafim, A.~T. Makariy, L.~Yong, A.~L. Thomas, E.~S.
  Warren, M.~Vivek, M.~Saurabh, J.~H. Peter, and P.~Ruslan.
\newblock Energy gap evolution across the superconductivity dome in single
  crystals of {(Ba$_{1-x}$K$_{x}$)Fe$_{2}$As$_{2}$}.
\newblock {\em Sci. Adv.}, 2:e1600807, 2016.

\bibitem{MZolliker2011HKawano-Furukawa}
H.~Kawano-Furukawa, C.~J. Bowell, J.~S. White, R.~W. Heslop, A.~S. Cameron,
  E.~M. Forgan, K.~Kihou, C.~H. Lee, A.~Iyo, H.~Eisaki, T.~Saito, H.~Fukazawa,
  Y.~Kohori, R.~Cubitt, C.~D. Dewhurst, J.~L. Gavilano, and M.~Zolliker.
\newblock Gap in {KFe$_{2}$As$_{2}$} studied by small-angle neutron scattering
  observations of the magnetic vortex lattice.
\newblock {\em Phys. Rev. B}, 84(2):024507, 2011.

\bibitem{CMeingast2016FHardy}
F.~Hardy, A.~E. Böhmer, L.~de' Medici, M.~Capone, G.~Giovannetti, R.~Eder,
  L.~Wang, M.~He, T.~Wolf, P.~Schweiss, R.~Heid, A.~Herbig, P.~Adelmann, R.~A.
  Fisher, and C.~Meingast.
\newblock Strong correlations, strong coupling, and s-wave superconductivity in
  hole-doped {BaFe$_{2}$As$_{2}$} single crystals.
\newblock {\em Phys. Rev. B}, 94(20):205113, 2016.

\bibitem{LTaillefer2013FFTafti}
F.~F. Tafti, A.~Juneau-Fecteau, M.~E. Delage, S.~Rene~de Cotret, J.~Ph Reid,
  A.~F. Wang, X.~G. Luo, X.~H. Chen, N.~Doiron-Leyraud, and Louis Taillefer.
\newblock Sudden reversal in the pressure dependence of {T$_c$} in the
  iron-based superconductor {KFe$_{2}$As$_{2}$}.
\newblock {\em Nat. Phys.}, 9(6):349--352, 2013.

\bibitem{RProzorov2014HKim}
H.~Kim, M.~A. Tanatar, Yong Liu, Zachary~Cole Sims, Chenglin Zhang, Pengcheng
  Dai, T.~A. Lograsso, and R.~Prozorov.
\newblock Evolution of {London} penetration depth with scattering in single
  crystals of
  {${\mathrm{K}}_{1\ensuremath{-}x}$${\mathrm{Na}}_{x}$${\mathrm{Fe}}_{2}$${\mathrm{As}}_{2}$}.
\newblock {\em Phys. Rev. B}, 89(17):174519, 2014.

\bibitem{CMeingast2014FHardy2}
F.~Hardy, R.~Eder, M.~Jackson, D.~Aoki, C.~Paulsen, T.~Wolf, P.~Burger,
  A.~Böhmer, P.~Schweiss, P.~Adelmann, R.~A. Fisher, and C.~Meingast.
\newblock Multiband superconductivity in {KFe$_{2}$As$_{2}$}: Evidence for one
  isotropic and several lilliputian energy gaps.
\newblock {\em J. Phys. Soc. Jpn.}, 83(1):014711, 2014.

\bibitem{HHKlauss2020VGrinenko}
V.~Grinenko, R.~Sarkar, K.~Kihou, C.~H. Lee, I.~Morozov, S.~Aswartham,
  B.~Buchner, P.~Chekhonin, W.~Skrotzki, K.~Nenkov, R.~Huhne, K.~Nielsch,
  S.~L. Drechsler, V.~L. Vadimov, M.~A. Silaev, P.~A. Volkov, I.~Eremin,
  H.~Luetkens, and H.~H. Klauss.
\newblock Superconductivity with broken time-reversal symmetry inside a
  superconducting s-wave state.
\newblock {\em Nat. Phys.}, 16(7):789--794, 2020.

\bibitem{JZhao2020SDShen}
S.~Shen, X.~Zhang, H.~Wo, Y.~Shen, Y.~Feng, A.~Schneidewind, P.~Cermak,
  W.~Wang, and J.~Zhao.
\newblock Neutron spin resonance in the heavily hole-doped {KFe$_{2}$As$_{2}$}
  superconductor.
\newblock {\em Phys. Rev. Lett.}, 124(1):017001, 2020.

\bibitem{HDing2009TSato}
T.~Sato, K.~Nakayama, Y.~Sekiba, P.~Richard, Y.~M. Xu, S.~Souma, T.~Takahashi,
  G.~F. Chen, J.~L. Luo, N.~L. Wang, and H.~Ding.
\newblock Band structure and {Fermi} surface of an extremely overdoped
  iron-based superconductor {KFe$_{2}$As$_{2}$}.
\newblock {\em Phys. Rev. Lett.}, 103(4):047002, 2009.

\bibitem{SUji2013TTerashima}
T.~Terashima, N.~Kurita, M.~Kimata, M.~Tomita, S.~Tsuchiya, M.~Imai, A.~Sato,
  K.~Kihou, C.~H. Lee, H.~Kito, H.~Eisaki, A.~Iyo, T.~Saito, H.~Fukazawa,
  Y.~Kohori, H.~Harima, and S.~Uji.
\newblock Fermi surface in {KFe$_{2}$As$_{2}$} determined via de {Haas-van
  Alphen} oscillation measurements.
\newblock {\em Phys. Rev. B}, 87(22):224512, 2013.

\bibitem{HHisatomo2014TYoshida}
T.~Yoshida, S.~I. Ideta, I.~Nishi, A.~Fujimori, M.~Yi, R.~G. Moore, S.~K. Mo,
  D.~Lu, Z.~X. Shen, Z.~Hussain, K.~Kihou, P.~M. Shirage, H.~Kito, C.~H. Lee,
  A.~Iyo, H.~Eisaki, and H.~Harima.
\newblock Orbital character and electron correlation effects on two- and
  three-dimensional {Fermi} surfaces in {KFe$_{2}$As$_{2}$} revealed by
  angle-resolved photoemission spectroscopy.
\newblock {\em Front. Phys.}, 2:17, 2014.

\bibitem{EHiroshi2010KKunihiro}
K.~Kihou, T.~Saito, S.~Ishida, M.~Nakajima, Y.~Tomioka, H.~Fukazawa, Y.~Kohori,
  T.~Ito, S.~Uchida, A.~Iyo, C.~H. Lee, and H.~Eisaki.
\newblock Single crystal growth and characterization of the iron-based
  superconductor {KFe$_2$As$_2$} synthesized by {KAs} flux method.
\newblock {\em J. Phys. Soc. Jpn.}, 79(12):124713, 2010.

\bibitem{UShinya2010TTaichi}
T.~Terashima, M.~Kimata, N.~Kurita, H.~Satsukawa, A.~Harada, K.~Hazama,
  M.~Imai, A.~Sato, K.~Kihou, C.~H. Lee, H.~Kito, H.~Eisaki, A.~Iyo, T.~Saito,
  H.~Fukazawa, Y.~Kohori, H.~Harima, and S.~Uji.
\newblock Fermi surface and mass enhancement in {KFe$_2$As$_2$} from de
  {Haas-van Alphen} effect measurements.
\newblock {\em J. Phys. Soc. Jpn.}, 79(5):053702, 2010.

\bibitem{XFWang2011JSKim}
J.~S. Kim, E.~G. Kim, G.~R. Stewart, X.~H. Chen, and X.~F. Wang.
\newblock Specific heat in {KFe${}_{2}$As${}_{2}$} in zero and applied magnetic
  field.
\newblock {\em Phys. Rev. B}, 83(17):172502, 2011.

\bibitem{LBoeri2012MAbdel-Hafiez}
M.~Abdel-Hafiez, S.~Aswartham, S.~Wurmehl, V.~Grinenko, C.~Hess, S.~L.
  Drechsler, S.~Johnston, A.~U.~B. Wolter, B.~Buchner, H.~Rosner, and
  L.~Boeri.
\newblock Specific heat and upper critical fields in {KFe${}_{2}$As${}_{2}$}
  single crystals.
\newblock {\em Phys. Rev. B}, 85(13):134533, 2012.

\bibitem{CMeingast2013FHardy}
F.~Hardy, A.~E. Bohmer, D.~Aoki, P.~Burger, T.~Wolf, P.~Schweiss, R.~Heid,
  P.~Adelmann, Y.~X. Yao, G.~Kotliar, J.~Schmalian, and C.~Meingast.
\newblock Evidence of strong correlations and coherence-incoherence crossover
  in the iron pnictide superconductor {KFe${}_{2}$As${}_{2}$}.
\newblock {\em Phys. Rev. Lett.}, 111(2):027002, 2013.

\bibitem{HVLohneysen2016FEilers}
F.~Eilers, K.~Grube, D.~A. Zocco, T.~Wolf, M.~Merz, P.~Schweiss, R.~Heid,
  R.~Eder, R.~Yu, J.~X. Zhu, Q.~Si, T.~Shibauchi, and H.~V. Lohneysen.
\newblock Strain-driven approach to quantum criticality in
  {AFe${}_{2}$As${}_{2}$} with {A=K, Rb, and Cs}.
\newblock {\em Phys. Rev. Lett.}, 116(23):237003, 2016.

\bibitem{XHChen2016YPWu}
Y.~P. Wu, D.~Zhao, A.~F. Wang, N.~Z. Wang, Z.~J. Xiang, X.~G. Luo, T.~Wu, and
  X.~H. Chen.
\newblock Emergent {Kondo} lattice behavior in iron-based superconductors
  {AFe${}_{2}$As${}_{2}$ (A=K, Rb, Cs)}.
\newblock {\em Phys. Rev. Lett.}, 116(14):147001, 2016.

\bibitem{MCapone2014Lde'Medici}
L.~de' Medici, G.~Giovannetti, and M.~Capone.
\newblock Selective {Mott} physics as a key to iron superconductors.
\newblock {\em Phys. Rev. Lett.}, 112(17):177001, 2014.

\bibitem{KTMoore2009}
K.~T. Moore and G.~van~der Laan.
\newblock Nature of the 5f states in actinide metals.
\newblock {\em Rev. Mod. Phys.}, 81(1):235--298, 2009.

\bibitem{CPfleiderer2009}
C.~Pfleiderer.
\newblock Superconducting phases of f-electron compounds.
\newblock {\em Rev. Mod. Phys.}, 81(4):1551--1624, 2009.

\bibitem{ZXShen2021}
J.~A. Sobota, Y.~He, and Z.~X. Shen.
\newblock Angle-resolved photoemission studies of quantum materials.
\newblock {\em Rev. Mod. Phys.}, 93(2):025006, 2021.

\bibitem{ZPYin2011}
Z.~P. Yin, K.~Haule, and G.~Kotliar.
\newblock Kinetic frustration and the nature of the magnetic and paramagnetic
  states in iron pnictides and iron chalcogenides.
\newblock {\em Nat. Mater.}, 10(12):932--935, 2011.

\bibitem{HHWen2015DLFang}
D.~Fang, X.~Shi, Z.~Du, P.~Richard, H.~Yang, X.~X. Wu, P.~Zhang, T.~Qian,
  X.~Ding, Z.~Wang, T.~K. Kim, M.~Hoesch, A.~Wang, X.~Chen, J.~Hu, H.~Ding, and
  H.~H. Wen.
\newblock Observation of a {van Hove} singularity and implication for
  strong-coupling induced {Cooper} pairing in {KFe$_2$As$_2$}.
\newblock {\em Phys. Rev. B}, 92(14):144513, 2015.

\bibitem{XJZhou2018}
X.~Zhou, S.~He, G.~Liu, L.~Zhao, L.~Yu, and W.~Zhang.
\newblock {New developments in laser-based photoemission spectroscopy and its
  scientific applications: a key issues review}.
\newblock {\em Rep. Prog. Phys.}, 81(6):062101, 2018.

\bibitem{XJZhou2021YQCai2}
Y.~Cai, T.~Xie, H.~Yang, D.~Wu, J.~Huang, W.~Hong, L.~Cao, C.~Liu, C.~Li,
  Y.~Xu, Q.~Gao, T.~Miao, G.~Liu, S.~Li, L.~Huang, H.~Luo, Z.~Xu, H.~Gao,
  L.~Zhao, and X.~J. Zhou.
\newblock Common ($\pi, \pi$) band folding and surface reconstruction in
  {FeAs}-based superconductors.
\newblock {\em Chin. Phys. Lett.}, 38(5):057404, 2021.

\bibitem{SShin2017TShimojima}
T.~Shimojima, W.~Malaeb, A.~Nakamura, T.~Kondo, K.~Kihou, C.~H. Lee, A.~Iyo,
  H.~Eisaki, S.~Ishida, M.~Nakajima, S.~I. Uchida, K.~Ohgushi, K.~Ishizaka, and
  S.~Shin.
\newblock Antiferroic electronic structure in the nonmagnetic superconducting
  state of the iron-based superconductors.
\newblock {\em Science Adv.}, 3(8):e1700466, 2017.

\bibitem{XJZhou2021YQCai}
Y.~Cai, J.~Huang, T.~Miao, D.~Wu, Q.~Gao, C.~Li, Y.~Xu, J.~Jia, Q.~Wang,
  Y.~Huang, G.~Liu, F.~Zhang, S.~Zhang, F.~Yang, Z.~Wang, Q.~Peng, Z.~Xu,
  L.~Zhao, and X.~Zhou.
\newblock Genuine electronic structure and superconducting gap structure in
  {(Ba$_{0.6}$K$_{0.4}$)Fe$_2$As$_2$} superconductor.
\newblock {\em Science Bull.}, 66(18):1839--1848, 2021.

\bibitem{JCCampuzano1998Norman}
M.~R. Norman, M.~Randeria, H.~Ding, and J.~C. Campuzano.
\newblock {Phenomenology of the low-energy spectral function in high-Tc
  superconductors}.
\newblock {\em Phys. Rev. B}, 57:4, 1998.

\bibitem{FBaumberger2010ATamai}
A.~Tamai, A.~Y. Ganin, E.~Rozbicki, J.~Bacsa, W.~Meevasana, P.~D. King,
  M.~Caffio, R.~Schaub, S.~Margadonna, K.~Prassides, M.~J. Rosseinsky, and
  F.~Baumberger.
\newblock Strong electron correlations in the normal state of the iron-based
  {FeSe$_{0.42}$Te$_{0.58}$} superconductor observed by angle-resolved
  photoemission spectroscopy.
\newblock {\em Phys. Rev. Lett.}, 104(9):097002, 2010.

\bibitem{MGreven2009GYu}
G.~Yu, Y.~Li, E.~M. Motoyama, and M.~Greven.
\newblock A universal relationship between magnetic resonance and
  superconducting gap in unconventional superconductors.
\newblock {\em Nat. Phys.}, 5(12):873--875, 2009.

\bibitem{DHLee2009FWang}
F.~Wang, H.~Zhai, Y.~Ran, A.~Vishwanath, and D.~H. Lee.
\newblock Functional renormalization-group study of the pairing symmetry and
  pairing mechanism of the {FeAs}-based high-temperature superconductor.
\newblock {\em Phys. Rev. Lett.}, 102(4):047005, 2009.

\bibitem{HTakagi2010THanaguri}
T.~Hanaguri, S.~Niitaka, K.~Kuroki, and H.~Takagi.
\newblock Unconventional s-wave superconductivity in {Fe(Se,Te)}.
\newblock {\em Science}, 328(5977):474--476, 2010.

\bibitem{KYamada2011CHLee}
C.~H. Lee, K.~Kihou, H.~Kawano-Furukawa, T.~Saito, A.~Iyo, H.~Eisaki,
  H.~Fukazawa, Y.~Kohori, K.~Suzuki, H.~Usui, K.~Kuroki, and K.~Yamada.
\newblock Incommensurate spin fluctuations in hole-overdoped superconductor
  {KFe$_{2}$As$_{2}$}.
\newblock {\em Phys. Rev. Letters}, 106(6):067003, 2011.

\bibitem{CHLee2016KHorigane}
K.~Horigane, K.~Kihou, K.~Fujita, R.~Kajimoto, K.~Ikeuchi, S.~Ji, J.~Akimitsu,
  and C.~H. Lee.
\newblock Spin excitations in hole-overdoped iron-based superconductors.
\newblock {\em Sci. Rep.}, 6:33303, 2016.

\bibitem{HEisaki2016KKihou}
K.~Kihou, T.~Saito, K.~Fujita, S.~Ishida, M.~Nakajima, K.~Horigane,
  H.~Fukazawa, Y.~Kohori, S.~Uchida, J.~Akimitsu, A.~Iyo, C.~H. Lee, and
  H.~Eisaki.
\newblock Single-crystal growth of {Ba$_{1-x}$K$_x$Fe$_2$As$_2$} by {KAs}
  self-flux method.
\newblock {\em J. Phys. Soc. Jpn.}, 85(3):034718, 2016.

\bibitem{JBardeen1957}
J.~Bardeen, L.~N. Cooper, and J.~R. Schrieffer.
\newblock Microscopic theory of superconductivity.
\newblock {\em Phys. Rev.}, 106(1):162--164, 1957.

\bibitem{JPHu2012HDing}
J.~Hu and H.~Ding.
\newblock Local antiferromagnetic exchange and collaborative fermi surface as
  key ingredients of high temperature superconductors.
\newblock {\em Sci. Rep.}, 2:381, 2012.

\end{thebibliography}

\noindent {\bf Acknowledgement}\\
 This work is supported by the National Key Research and Development Program of China (Grant No. 2021YFA1401800, 2017YFA0302900, 2018YFA0704200, 2018YFA0305600, 2019YFA0308000 and 2022YFA1604203), the National Natural Science Foundation of China (Grant No. 11888101, 11922414 and 11974404), the Strategic Priority Research Program (B) of the Chinese Academy of Sciences (Grant No. XDB25000000 and XDB33000000) , the Youth Innovation Promotion Association of CAS (Grant No. Y2021006 and Y202001) and Synergetic Extreme Condition User Facility (SECUF).

\vspace{3mm}
\noindent {\bf Author Contributions}\\
 X.J.Z., L.Z. and D.S.W. proposed and designed the research. D.S.W, W.S.H., Y.J.S, H.Q.L and S.L.L. contributed in sample growth and the magnetic and resistivity measurements. J.J.J., J.G.Y., H.T.Y, H.T.R, P.A., X.Z., C.H.Y., T.M.M., C.L.L., S.J.Z., F.F.Z., F.Y., Z.M.W., N.Z., L.J.L., R.K.L., X.Y.W., Q.J.P., H.Q.M., G.D.L., Z.Y.X., L.Z. and X.J.Z. contributed to the development and maintenance of the ARPES systems and related software development. D.S.W. carried out the ARPES experiment with J.J.J. and J.G.Y.. D.S.W., L.Z. and X.J.Z. analyzed the data. X.X.W. contributed to the band structure calculations. X.J.Z., L.Z. and D.S.W. wrote the paper. All authors participated in discussion and comment on the paper.

\vspace{3mm}
\noindent {\bf Competing Interests Statement}\\
The authors declare that they have no competing financial interests.
\vspace{3mm}

\noindent{\bf Additional information}\\
Supplementary information is available in the online version of the paper.
Correspondence and requests for materials should be addressed to L. Zhao and X. J. Zhou.

\newpage

\end{document}



\title{{\rm\it{Supplementary Materials for}}\\
Nodal $\textsl{s}_\pm$ Pairing Symmetry in an Iron-Based Superconductor with only Hole Pockets\\}

\author{Dingsong Wu$^{1,2,\dagger}$, Junjie Jia$^{1,2,\dagger}$, Jiangang Yang$^{1,2,\dagger}$, Wenshan Hong$^{1,2}$, Yingjie Shu$^{1,2}$, Taimin Miao$^{1,2}$, Hongtao Yan$^{1,2}$, Hongtao Rong$^{1,2}$, Ping Ai$^{1,2}$, Xing Zhang$^{1,2}$, Chaohui Yin$^{1,2}$, Chenlong Li$^{3}$, Shenjin Zhang$^{3}$, Fengfeng Zhang$^{3}$, Feng Yang$^{3}$, Zhimin Wang$^{3}$, Nan Zong$^{3}$, Lijuan Liu$^{3}$, Rukang Li$^{3}$, Xiaoyang Wang$^{3}$, Qinjun Peng$^{3}$, Hanqing Mao$^{1,2,5}$, Guodong Liu$^{1,2,5}$, Shiliang Li$^{1,2,5}$, Huiqian Luo$^{1,2,5}$, Xianxin Wu$^{4}$, Zuyan Xu$^{3}$, Lin Zhao$^{1,2,5\bigstar}$ and X. J. Zhou$^{1,2,5\bigstar}$}

\affiliation{
\\$^{1}$Beijing National Laboratory for Condensed Matter Physics, Institute of Physics, Chinese Academy of Sciences, Beijing 100190, China.
\\$^{2}$University of Chinese Academy of Sciences, Beijing 100049, China.
\\$^{3}$Technical Institute of Physics and Chemistry, Chinese Academy of Sciences, Beijing 100190, China.
\\$^{4}$Institute of Theoretical Physics, Chinese Academy of Sciences, Beijing 100190, China.
\\$^{5}$Songshan Lake Materials Laboratory, Dongguan, Guangdong 523808, China.
}
\maketitle
\begin{figure*}[tbp]
\begin{center}
\includegraphics[width=1\textwidth,angle=0]{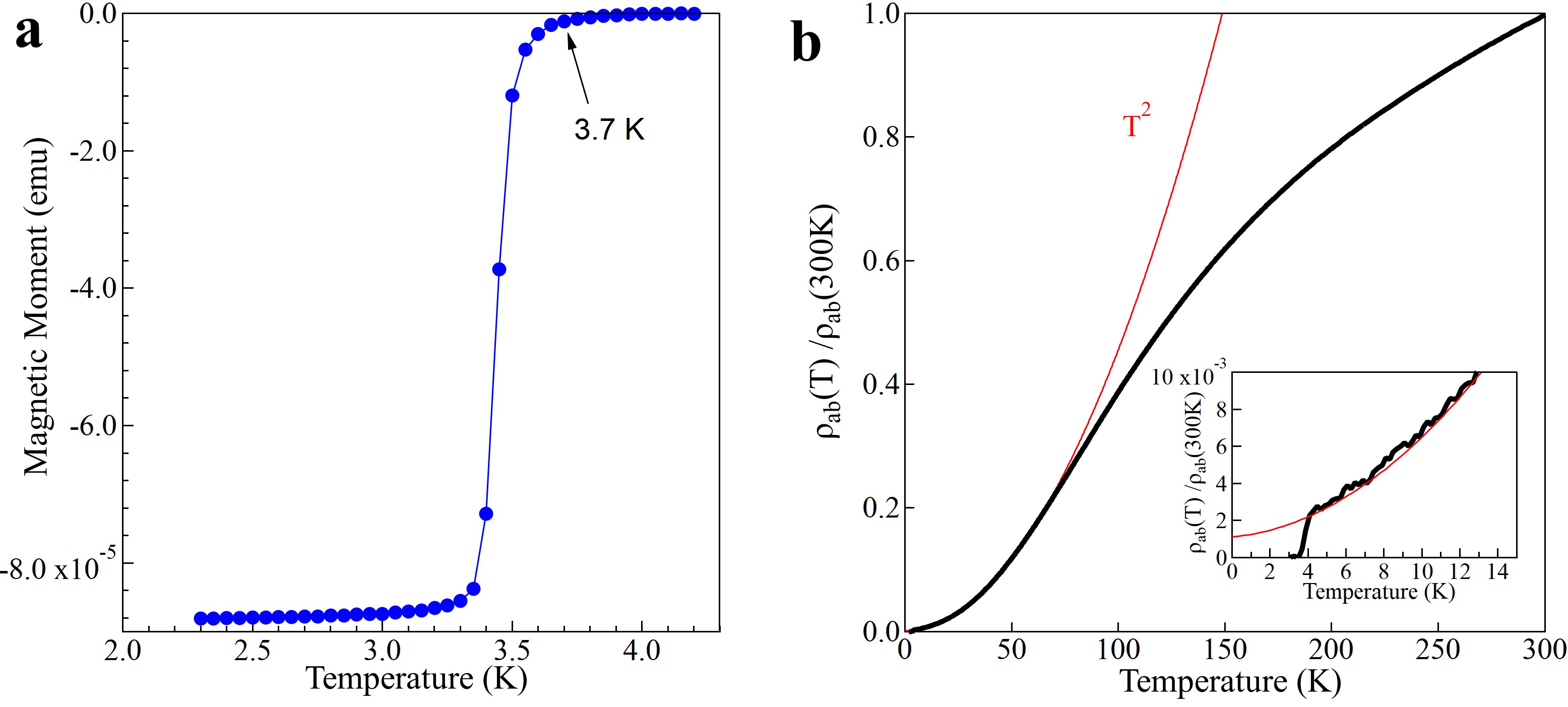}
\end{center}
\caption{{\bf Magnetic and transport measurements of the KFe$_2$As$_2$ samples.} \textbf{a}, Magnetic susceptibility measured under a magnetic field of 2\,Oe. The measured T$_C$ (onset) is 3.7\,K and the transition width (10$\%$-90$\%$) is $\sim$0.2\,K. \textbf{b}, In-plane resistivity normalized by the value at 300\,K. The low temperature part is fitted by a+bT$^2$ (red line) and the fitted a\,=\,0.0011. The inset shows the data near the superconducting transition. Considering the resistivity at 300\,K and the extracted value at zero temperature, a RRR (residual resistivity ratio) value of 910 is obtained for the measured KFe$_2$As$_2$ sample.
   }
\label{sample}
\end{figure*}

\noindent\textbf{1. Characterization of KFe$_2$As$_2$ Single Crystals}\vspace{3mm}\\
\hspace*{6mm}Figure \ref{sample} shows the magnetic and transport measurement results of the KFe$_2$As$_2$ single crystal samples we used in our ARPES measurements. The magnetic susceptibility data exhibits a narrow superconducting transition with a width (10$\%$-90$\%$) of $\sim$0.2\,K and an onset T$_C$ of 3.7\,K (Fig. \ref{sample}a). The low temperature part of the resistivity can be fitted by a+bT$^2$ (red line in Fig. \ref{sample}b) and the fitted a\,=\,0.0011.  Considering the resistivity at 300\,K and the extracted value at zero temperature, a RRR (residual resistivity ratio) value of 910 is obtained for the measured KFe$_2$As$_2$ samples. The relatively high T$_C$, the sharp superconducting transition and the large RRR value indicate high quality of our single crystal samples\cite{LTaillefer2012JPhReid}. \\

\begin{figure*}[tbp]
\begin{center}
\includegraphics[width=1\textwidth,angle=0]{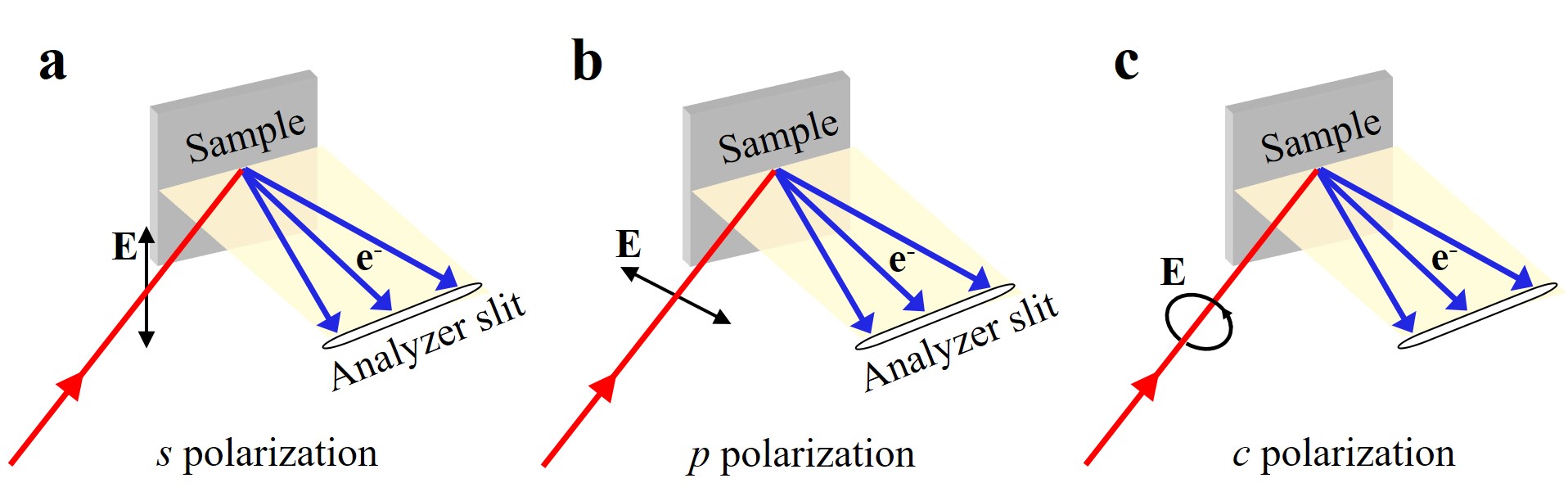}
\end{center}
\caption{{\bf Schematic illustration of the polarization geometries used in the ARPES measurements.} \textbf{a}, s polarization geometry where the electric field vector \textbf{E} is perpendicular to the plane formed by the incident laser light and the analyzer slit. \textbf{b}, p polarization geometry where the electric field vector \textbf{E} is in the plane formed by the incident laser light and the analyzer slit. \textbf{c}, c polarization geometry where the electric field vector \textbf{E} is circular.
   }
\label{polar}
\end{figure*}

\noindent\textbf{2. Different Polarization Geometries Used in the ARPES Measurements}\vspace{3mm}\\
\hspace*{6mm}The ARPES measurements involve photoemission matrix element effects where the observation of bands depends on the used polarization geometries\cite{ZXShen2021}. In order to get the complete electronic structure and the related superconducting gap structure, different polarization geometries need to be used in the ARPES measurements. In our present experiments, three kinds of light polarizations are used as shown in Fig. \ref{polar} which include s polarization (Fig. \ref{polar}a), p polarization (Fig. \ref{polar}b) and c polarization (Fig. \ref{polar}c). \\

\begin{figure*}[tbp]
\begin{center}
\includegraphics[width=1\textwidth,angle=0]{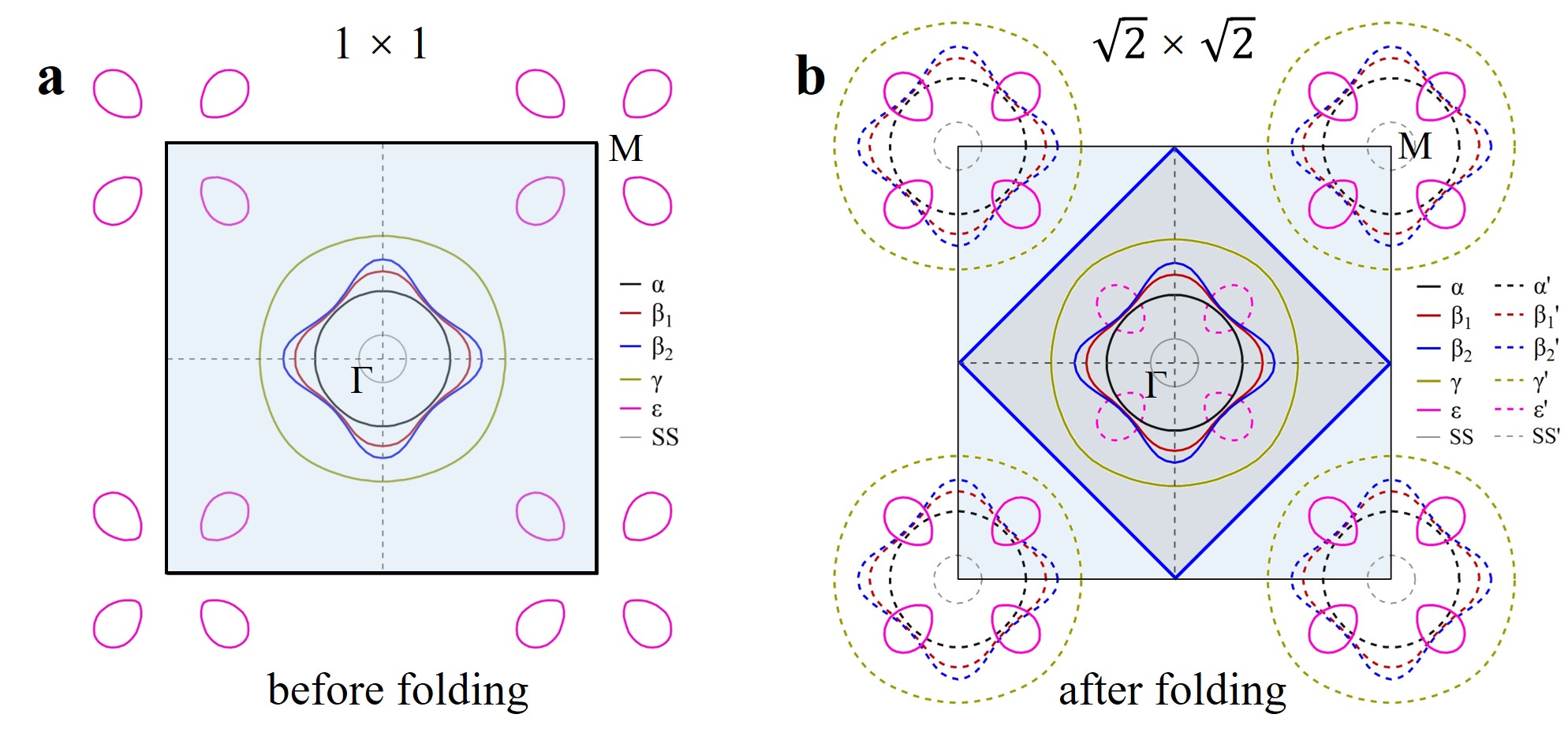}
\end{center}
\caption{{\bf Band folding in KFe$_2$As$_2$ caused by the $\sqrt{2}$\,$\times$\,$\sqrt{2}$ surface reconstruction.} \textbf{a}, Fermi surface before folding. \textbf{b}, Fermi surface after folding due to the $\sqrt{2}$\,$\times$\,$\sqrt{2}$ surface reconstruction. The first Brillouin zone shrinks by a half from the initial one (thick black lines) in \textbf{a} to the new reconstructed one (thick blue lines) in \textbf{b}. The Fermi surface are folded between $\Gamma$ and M points. Particularly around $\Gamma$, in addition to the initial $\alpha$, $\beta$ and SS Fermi surface, the four hole pockets around M are folded to the $\Gamma$ point forming four $\varepsilon'$ Fermi surface.
   }
\label{folding}
\end{figure*}

\noindent\textbf{3. Surface Reconstruction and Band Folding in KFe$_2$As$_2$}\vspace{3mm}\\
\hspace*{6mm}KFe$_2$As$_2$ has a $\sqrt{2}$\,$\times$\,$\sqrt{2}$ surface reconstruction, as observed in the previous STM measurements\cite{HHWen2015DLFang}. With this reconstruction, the electronic structure around the $\Gamma$ point and the M point will be folded to each other, as shown in Fig. \ref{folding}. In (Ba$_{1-x}$K$_x$)Fe$_2$As$_2$ system, due to such a $\sqrt{2}$\,$\times$\,$\sqrt{2}$ surface reconstruction, the band folding between $\Gamma$ and M points is commonly observed \cite{SShin2017TShimojima, XJZhou2021YQCai2, XJZhou2021YQCai}. Particularly, in addition to the initial $\alpha$, $\beta$, $\gamma$ and surface state (SS) Fermi surface, the four hole pockets around M are folded to the $\Gamma$ point forming four $\varepsilon'$ Fermi surface (Fig. \ref{folding}b). This makes it possible to measure the complete electronic structure and superconducting gap structure of KFe$_2$As$_2$ by using our laser ARPES.\\

\begin{figure*}[tbp]
\begin{center}
\includegraphics[width=1\textwidth,angle=0]{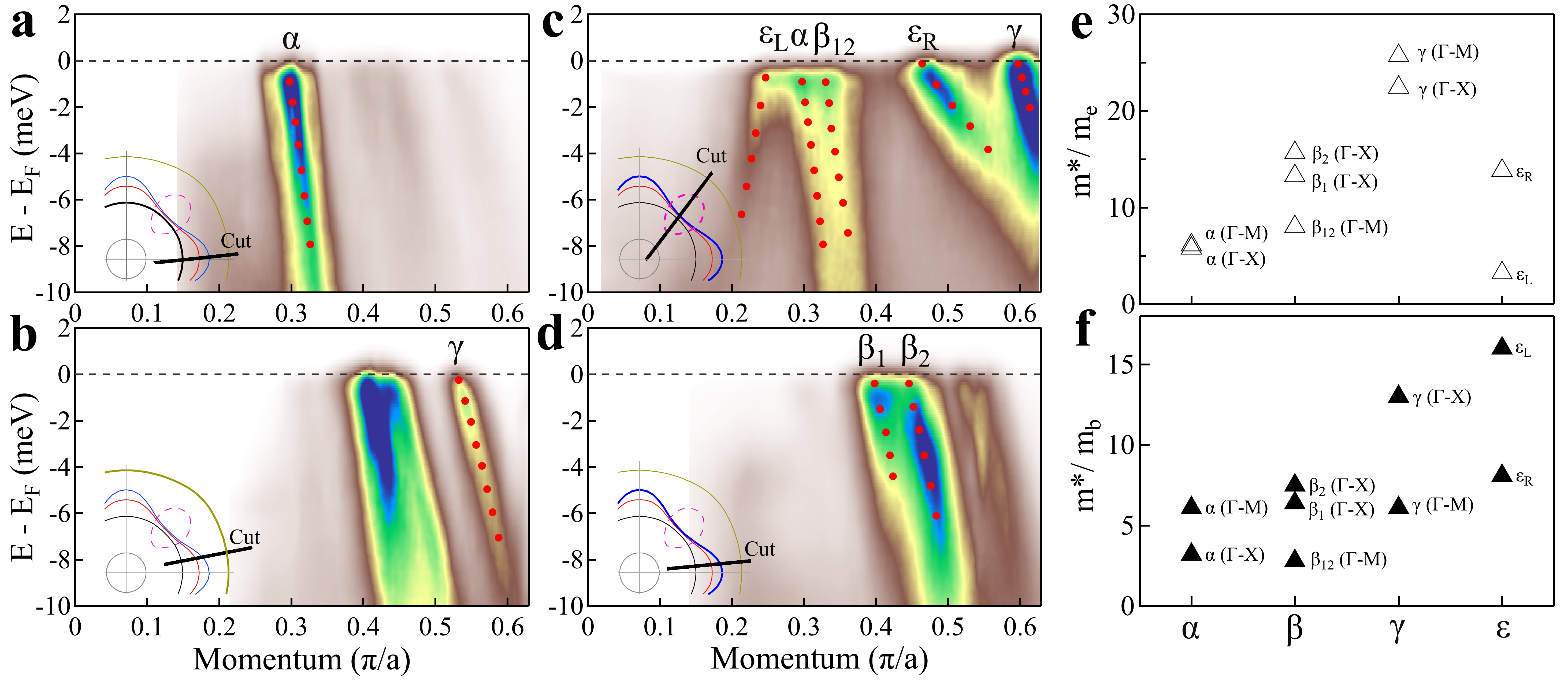}
\end{center}
\caption{{\bf Determination of the effective mass for different bands in KFe$_2$As$_2$.} \textbf{a-d}, Band structures measured along the momentum cuts marked by the black lines in each inset. The dispersion of the bands is marked by red dotted lines. \textbf{e}, The extracted effective mass of different bands in the unit of free electron mass m$_e$. It is estimated from m$^*$ = ${\hbar}$k$_F$/v$_F$ where the Fermi velocity v$_F$ is obtained from the dispersions in \textbf{a-d}. \textbf{f}, The extracted effective mass of different bands in the unit of the band electron mass m$_b$ which is obtained from the calculated band structures.
   }
\label{effeMass}
\end{figure*}

\noindent\textbf{4. Determination of the Effective Mass for Different Fermi Surface}\vspace{3mm}\\
\hspace*{6mm} In Fig. \ref{effeMass}, we extracted the effective mass from m$^*$ = ${\hbar}$k$_F$/v$_F$, with the Fermi momentum k$_F$ and the Fermi velocity v$_F$ extracted from the measured band dispersions (red lines) in Fig. \ref{effeMass}a-d. We show m$^*$/m$_e$ in Fig. \ref{effeMass}e and m$^*$/m$_b$ in Fig. \ref{effeMass}f, with m$_e$ and m$_b$ being the free electron mass and the band electron mass, respectively. The measured effective mass is also listed in Table \ref{effeMass2}. All the bands exhibit strongly enhanced effective mass with the m$^*$/m$_e$ of the $\gamma$ band reaching over 20 (Fig. \ref{effeMass}e). When compared with the calculated band structures, the $\varepsilon$ band shows the strongest band renormalization with m$^*$/m$_b$ over 15 (Fig. \ref{effeMass}f).

We compare our measured Fermi surface area, effective mass m$^*$ and Sommerfeld coefficient $\gamma$ of KFe$_2$As$_2$ with the previous ARPES measurements\cite{HHisatomo2014TYoshida}, dHvA measurements\cite{SUji2013TTerashima}, previous band structure calculations\cite{UShinya2010TTaichi,HHisatomo2014TYoshida} and our band structure calculations in Table \ref{effeMass2}. The areas we measured for each Fermi surface are very close to those from the dHvA measurements\cite{UShinya2010TTaichi, SUji2013TTerashima}. The overall doping level is consistent with 0.5\,hole/Fe that is expected in KFe$_2$As$_2$. With the effective mass in Fig. \ref{effeMass} and Table \ref{effeMass2}, we can calculate the Sommerfeld coefficient $\gamma$ contributed from each Fermi surface sheet by using the relation $\gamma$ = 1.452\,$\mu_0^ *$\,(mJ\,mol$^{-1}$\,K$^{-2}$) where $\mu_0^ *$ = m$^*$/m$_e$\cite{SUji2013TTerashima}. The total Sommerfeld coefficient is obtained by summing up the contributions from all the Fermi surface sheets, and the obtained value of 109.6\,mJ\,mol$^{-1}$\,K$^{-2}$ is very close to that from the specific heat measurements with $\gamma$ = 101.7\,mJ\,mol$^{-1}$\,K$^{-2}$\,\cite{XFWang2011JSKim} or 103\,mJ\,mol$^{-1}$\,K$^{-2}$\,\cite{CMeingast2013FHardy}. The four $\varepsilon$ hole pockets contribute nearly half of the total Sommerfeld coefficient, indicating their important role in determining the macroscopic physical properties.\\

\noindent\textbf{5. Examination on Various Methods in Extracting the Superconducting Gap from ARPES Measurements}\vspace{3mm}\\
\hspace*{6mm} It is important to examine carefully on various ways to extract the superconducting gap from the ARPES measurements. For the angle-integrated data, the Dynes function is usually used\cite{RCDynes1978}:
\begin{equation}\label{Dynes}
\rho(\omega,\Gamma) = \dfrac{\omega-i\Gamma}{[(\omega-i\Gamma)^2-\Delta^2]^{1/2}}
\end{equation}
\noindent in which $\Gamma$ is the scattering rate and $\Delta$ is the superconducting gap. Such a method has been used in angle integrated photoemission measurements of the conventional superconductors Pb and Nb\cite{SShin2000AChainani} and heavy Fermion superconductor CeRu$_2$\cite{SWatanabe2005TKiss}. For the angle-resolved photoemission measurements, the superconducting gap is usually obtained from the photoemission spectrum (EDC) at the Fermi momentum. In this case, there are two different ways that are often used to fit the measured spectrum for getting the superconducting gap. The first is to use the BCS formula\cite{JBardeen1957}:
\begin{equation}\label{BCSEquation}
A_{BCS}(k,\omega) = \dfrac{1}{\pi}\{ \dfrac{|u_k|^2 \Gamma}{(\omega-E_k)^2+\Gamma^2} + \dfrac{|v_k|^2 \Gamma}{(\omega+E_k)^2+\Gamma^2}\}
\end{equation}
\noindent in which $|u_k|^2=1-|v_k|^2=\dfrac{1}{2}(1+\dfrac{\epsilon_k}{E_k})$ where $\epsilon_k$ represents the band in the normal state and $E_k$ represents the band in the superconducting state with $E_k=\sqrt{\epsilon_k^2+\Delta_k^2}$. At the Fermi momentum k$_F$, the formula reduces into
\begin{equation}\label{BCSEquation2}
A_{BCS}(k_F,\omega) = \dfrac{1}{2\pi}\{ \dfrac{\Gamma}{(\omega-\Delta)^2+\Gamma^2} + \dfrac{\Gamma}{(\omega+\Delta)^2+\Gamma^2}\}
\end{equation}
\noindent The second way to extract the superconducting gap from the measured EDC at k$_F$ is to fit the corresponding symmetrized EDC by the Norman function\cite{Norman1998}:
\begin{equation}\label{NormanFunction}
A(k,\omega) = \dfrac{1}{\pi}\dfrac{\Sigma''(k,\omega)}{(\omega-\epsilon_k-\Sigma'(k,\omega))^2 + \Sigma''(k,\omega)^2}\\
\end{equation}
\begin{equation}\label{NormanFunction}
\Sigma(k,\omega) = \Sigma'(k,\omega) + i\Sigma''(k,\omega) = -i\Gamma_1 + \dfrac{\Delta^2}{\omega + \epsilon(k)}
\end{equation}
\noindent In real practice, in fitting the measured photoemission spectrum, the background and the instrumental resolution (momentum resolution and energy resolution in ARPES measurements) also need to be considered.

\begin{figure*}[tbp]
\begin{center}
\includegraphics[width=1\textwidth,angle=0]{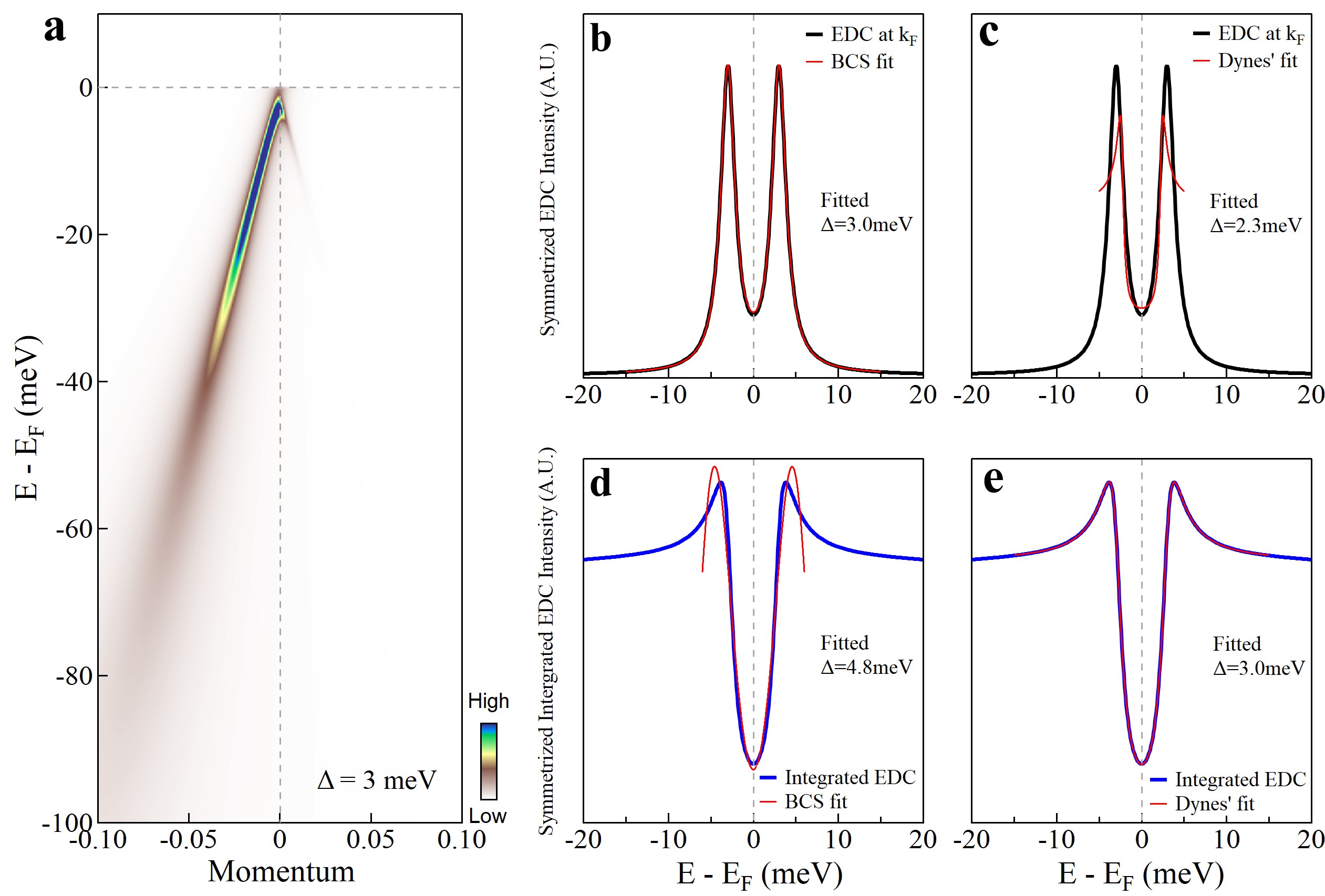}
\end{center}
\caption{{\bf Examination on different methods in extracting the superconducting gap.} \textbf{a}, Simulated band structure in the superconducting state using the BCS formula.  A linear bare band and the electron self energy of the Fermi liquid form are used. A superconducting gap $\Delta$ = 3\,meV and an electron impurity scattering of $\Gamma$ = 1\,meV are put in the simulation. \textbf{b}, Symmetrized EDC at the Fermi momentum (black line) fitted by the BCS formula (red line). The EDC can be well fitted with the fitted gap size in a good agreement with the put-in value. \textbf{c}, Symmetrized EDC at the Fermi momentum (black line) fitted by the Dynes function (red line). The EDC cannot be fitted well; the fitted gap size deviates from the value that is put in. \textbf{d}, Symmetrized EDC obtained by integrating over the entire momentum space in \textbf{a} (blue line). It is fitted by the BCS formula (red line). The integrated EDC cannot be fitted well and the fitted gap size is different from the gap value that is put in. \textbf{e}, The same symmetrized integrated EDC (blue line) fitted by the Dynes function (red line). The integrated EDC can be well fitted and the fitted gap size is consistent with the gap value that is put in.
   }
\label{BCSandDynes}
\end{figure*}

 The Dynes function fitting works for the angle-integrated photoemission spectrum but not for the single EDC at the Fermi momentum. To illustrate this point, we simulated the single particle spectral function in the superconducting state using the BCS formula, as shown in Fig. \ref{BCSandDynes}a. Here a linear bare band and the electron self energy of the Fermi liquid form are used. Fig. \ref{BCSandDynes}b shows the symmetrized EDC at the Fermi momentum. It exhibits sharp EDC peaks and can be well fitted by the BCS formula. When we try to fit the curve by the Dynes function, as shown in Fig. \ref{BCSandDynes}c, it cannot be fitted at all because the symmetrized EDC (black curve) and the Dynes function (red curve) have totally different lineshape. This indicates that it is not appropriate to fit the single EDC at the Fermi momentum by the Dynes function in order to get the superconducting gap. On the other hand, for the momentum integrated EDC, it cannot be fitted well by the BCS formula, as shown in Fig. \ref{BCSandDynes}d. But it can be well fitted by the Dynes function (Fig. \ref{BCSandDynes}e).

\begin{figure*}[tbp]
\begin{center}
\includegraphics[width=1\textwidth,angle=0]{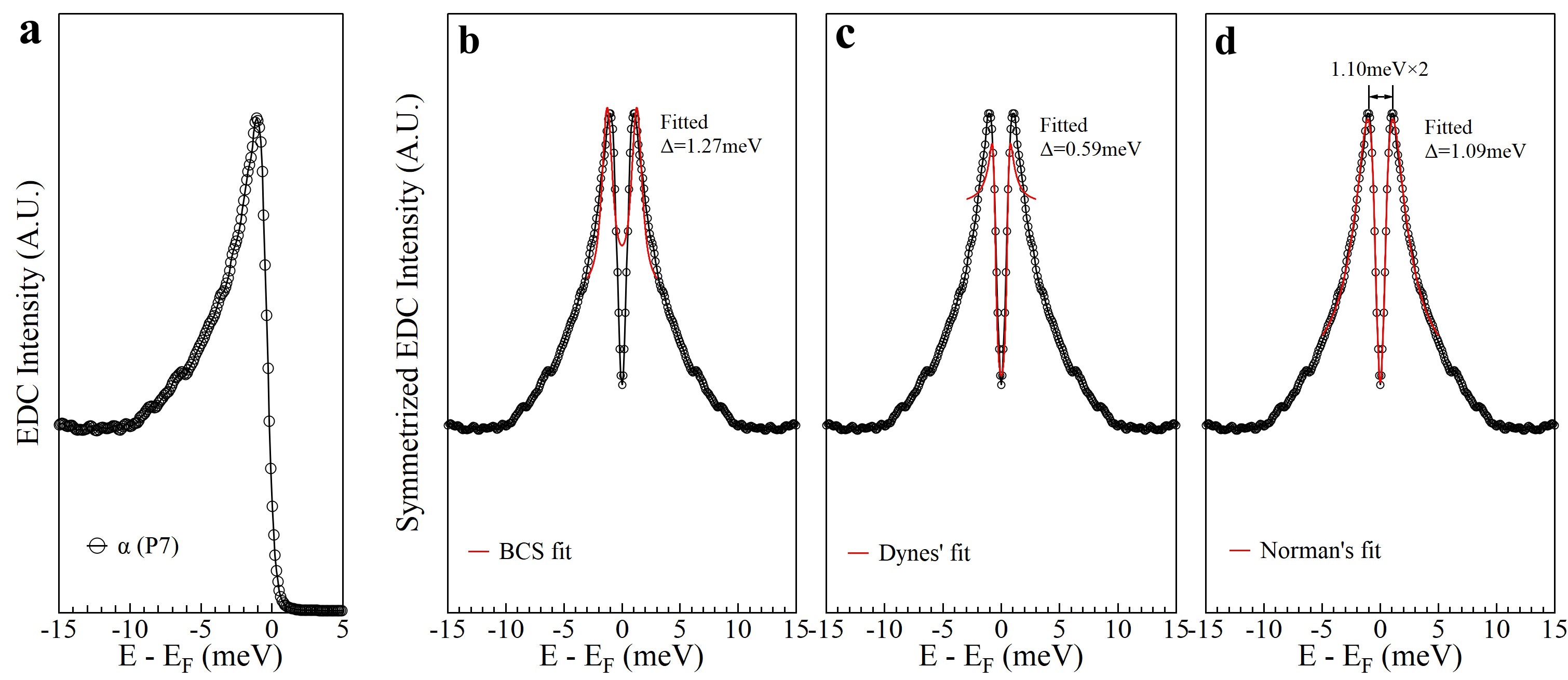}
\end{center}
\caption{{\bf Extraction of the superconducting gap from our data using different methods.} \textbf{a}, An EDC from our data which is measured at the momentum position P7 along the $\alpha$ Fermi surface in Fig. \ref{alpha}. \textbf{b}, Symmetrized EDC obtained from \textbf{a}. It is fitted by the BCS formula (red line). The symmetrized EDC cannot be well fitted. \textbf{c}, The same symmetrized EDC fitted by the Dynes function (red line). The fitted curve deviates strongly from the measured symmetrized EDC. \textbf{d}, The same symmetrized EDC fitted by the Norman's function (red line). The symmetrized EDC can be well fitted with the fitted gap size of 1.09\,meV. It is consistent with the gap value obtained from the energy difference between the two peaks (2$\Delta$).
   }
\label{EDCfit}
\end{figure*}

Fig. \ref{EDCfit}a shows a typical EDC we measured at the Fermi momentum of the $\alpha$ band at 0.9\,K in the superconducting state of KFe$_2$As$_2$. Here a sharp EDC peak is observed which is quite different from the previous ARPES measurements where basically no EDC peaks are observed\cite{SShin2012KOkazaki}. We find that, when there is a sharp peak, there is no way that the Dynes function can fit the measured spectrum (Fig. \ref{EDCfit}c). The symmetrized EDC can be partially fitted by the BCS formula but the fitting is not quite satisfactory (Fig. \ref{EDCfit}b). In this case, the Norman function can provide a best fitting of the symmetrized EDC, as seen in Fig. \ref{EDCfit}d. We note that, in the Norman function fitting, the energy position of the EDC peaks directly correspond to the superconducting gap size.

\begin{figure*}[tbp]
\begin{center}
\includegraphics[width=0.8\textwidth,angle=0]{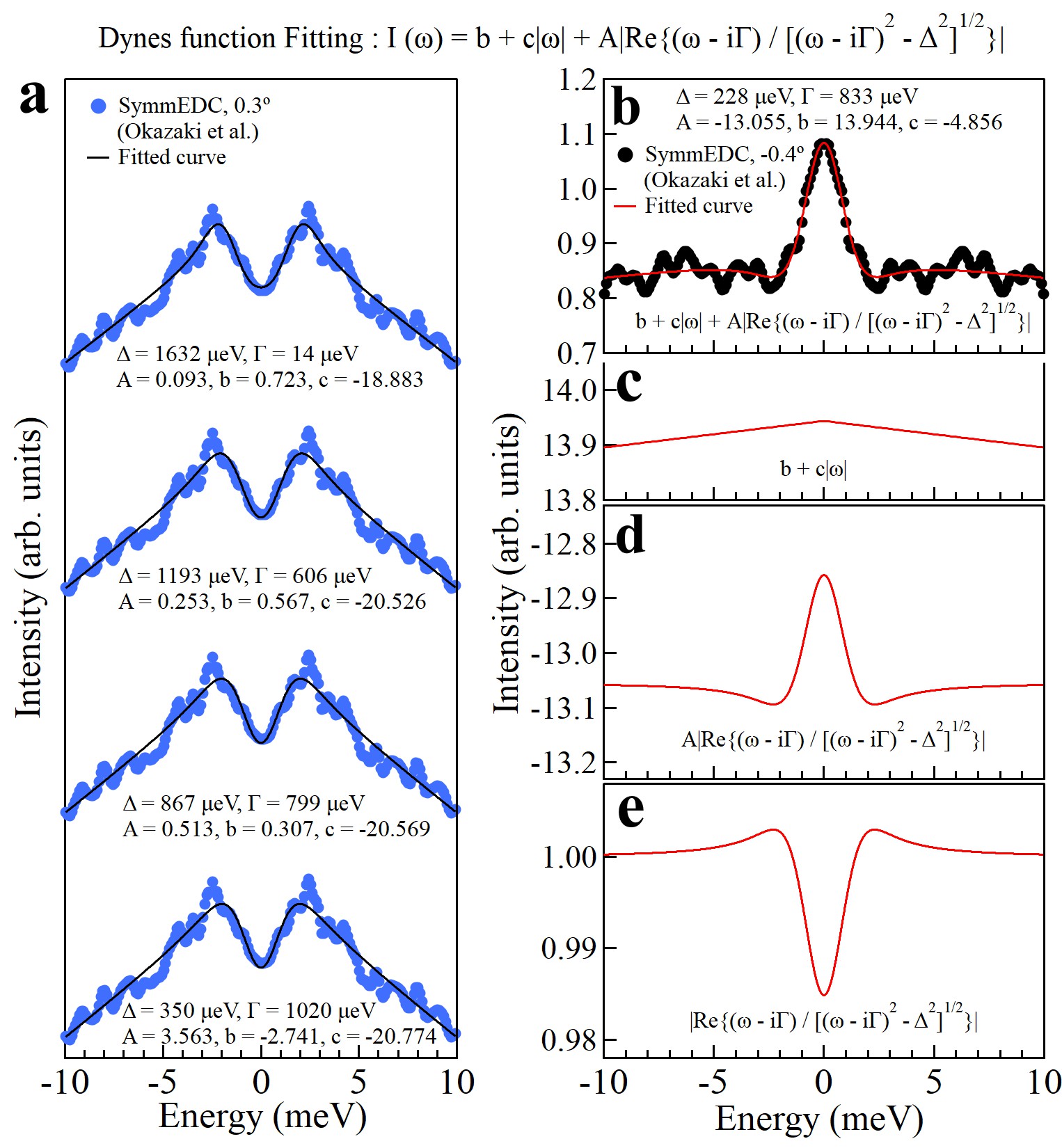}
\end{center}
\caption{{\bf Examination on the fitting procedures to extract the superconducting gap used in ref. \onlinecite{SShin2012KOkazaki}.} \textbf{a}, A symmetrized EDC fitted by the Dynes function. The symmetrized EDC (blue dots) is reproduced from Fig. S8 in the Supplementary Material of ref. \onlinecite{SShin2012KOkazaki}. It is fitted by the Dynes function considering a linear background and the convolution of the energy resolution (1.2\,meV); the fitting formula is shown on the top of the figure. The fitting depends sensitively on the initial values of the fitting parameters and different groups of the fitted parameters can be obtained. Here four groups of the fitted results are shown and the extracted gap size may vary from 350\,$\mu$eV (bottom curve) to 1632\,$\mu$eV (top curve). \textbf{b}, The invalid fitting of a peak in the symmetrized EDC by the Dynes function. The symmetrized EDC (black dots) is reproduced from Fig. 2E for the Fermi surface angle -0.4$^\circ$ in ref. \onlinecite{SShin2012KOkazaki}. It is fitted by the Dynes function plus a linear background with the fitting formula shown on the top of the figure. After a careful decomposition, it is found that the fitted curve (red line in \textbf{b}) is obtained from the initial intrinsic Dynes function (\textbf{e}), multiplied by a negative A to flip its shape (\textbf{d}), and then added on the top of a large positive background (\textbf{c}). The Dynes function cannot produce a peak at the Fermi level. This fitting is invalid because the negative Dynes function term and the large background term are both unphysical.
   }
\label{EDCShin}
\end{figure*}

Now we give a brief discussion on the superconducting gap results in the previous ARPES measurements of KFe$_2$As$_2$\cite{SShin2012KOkazaki}. In addition to the issues that the Fermi surface and the band structures were not clearly resolved and no sharp EDC peaks in the superconducting state were observed, there are serious problems in their procedure to extract the superconducting gap. First, they used the Dynes function to fit the single EDC at the Fermi momentum. As we have discussed before, the Dynes function is not applicable to fitting the EDC at the Fermi momentum even though the EDC does not show a sharp peak. Second, the Dynes function fitting does not provide the high precision of the superconducting gap on the order of $\mu$eV as they claimed. In Fig. \ref{EDCShin}a we replot a symmetrized EDC from their measurement and fit it by the Dynes function. We find that the fitting depends sensitively on the initial values of the fitting parameters and different groups of the fitted parameters can be obtained. In the figure we show four groups of the fitted results as examples. The extracted gap size may vary from 350\,$\mu$eV (bottom curve in Fig. \ref{EDCShin}a) to 1632\,$\mu$eV (top curve in Fig. \ref{EDCShin}a). This indicates that there is a considerable uncertainty in their determination of the gap size. Third, there is a mistake in their fitting of the peak structure in the symmetrized EDCs. When there is a peak at the Fermi level in the symmetrized EDC, they could still fit it by the Dynes function and extract the superconducting gap. However, it is not possible for the Dynes function to produce a peak at the Fermi level, as shown in Fig. \ref{BCSandDynes}e. To check on the possible problem in their fitting process\cite{SShin2012KOkazaki}, we also fitted their data by the Dynes function (Fig. \ref{EDCShin}b) and decomposed the fitted result into separate components (Fig. \ref{EDCShin}c-e). It turns out that their fitted curve (red line in Fig. \ref{EDCShin}b) is obtained from the initial intrinsic Dynes function (Fig. \ref{EDCShin}e), multiplied by a negative A to flip its shape (Fig. \ref{EDCShin}d), and then added on top of a large positive background (Fig. \ref{EDCShin}c). This fitting is invalid because the negative Dynes function term and the large background term are both unphysical. Therefore, there are serious problems in their superconducting gap determination and the fine structures of the superconducting gap, including the gap structure along the $\gamma$ Fermi surface and the `octet-line node structure' along the $\beta$ Fermi surface, are not reliable.\\

\begin{figure*}[tbp]
\begin{center}
\includegraphics[width=1\textwidth,angle=0]{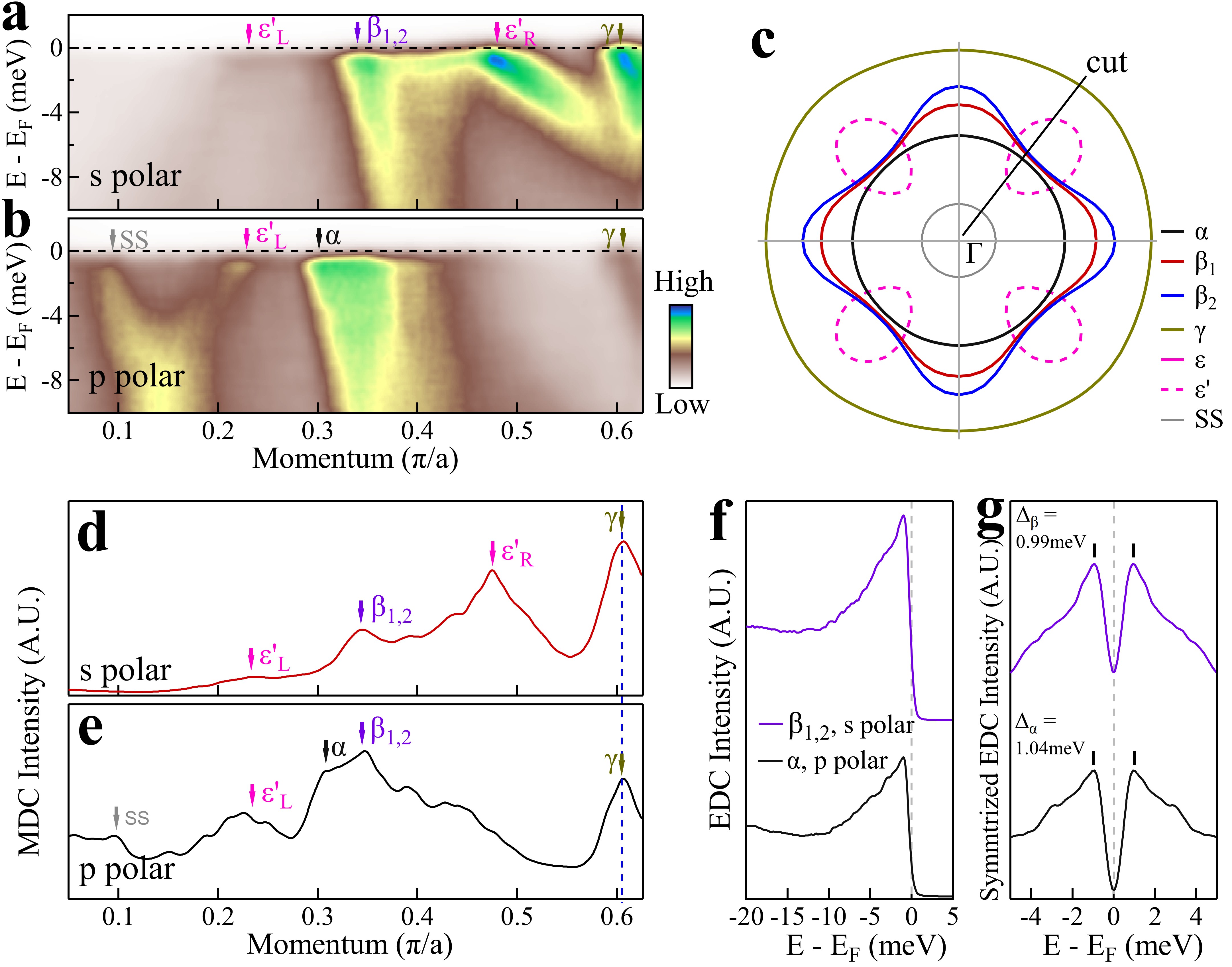}
\end{center}
\caption{{\bf Differentiation between the $\alpha$ and $\beta$ bands by using different polarization geometries.} \textbf{a}, Band structure along the momentum cut measured under the s polarization geometry. The location of the momentum cut is marked by a black line in \textbf{c}. In this case, the $\beta_{1,2}$, $\varepsilon'_R$ and $\gamma$ bands are clearly observed while the $\alpha$, $\varepsilon'_L$ and SS bands are suppressed. \textbf{b}, Band structure along the same momentum cut measured under the p polarization geometry. In this case, the $\alpha$, $\beta_{1,2}$, $\varepsilon'_L$ and SS bands are clearly observed while the $\varepsilon'_R$ and $\gamma$ bands are suppressed. \textbf{d-e}, The momentum distribution curves (MDCs) at the Fermi level obtained from \textbf{a} and \textbf{b}. \textbf{f}, EDCs at the Fermi momentum $\beta_{1,2}$ in \textbf{a} (purple line) and at the Fermi momentum $\alpha$ in \textbf{b} (black line). \textbf{g}, Corresponding symmetrized EDCs obtained from \textbf{f}. The extracted superconducting gap of the $\alpha$ and $\beta_{1,2}$ bands is 1.04 and 0.99\,meV, respectively.
   }
\label{alphabeta}
\end{figure*}

\noindent\textbf{6. Differentiation between the $\alpha$ and $\beta$ Bands by Using Different Polarization Geometries}\vspace{3mm}\\
\hspace*{6mm}Along the $\Gamma$-M direction, the $\alpha$ band and the $\beta$ band are close to each other, as shown in Fig. 1b. To distinguish these two bands, we carried out the measurements along the $\Gamma$-M direction under two different polarization geometries, as shown in Fig. \ref{alphabeta}. The $\beta$ band can be observed in both s and p polarizations while the $\alpha$ band can be observed only in the p polarization (Fig. \ref{alphabeta}a, b, d and e). These two bands can then be disentangled and we can obtain the EDCs at the respective k$_F$s of  the $\alpha$ and $\beta$ bands (Fig. \ref{alphabeta}f). The extracted superconducting gap of the $\alpha$ band (1.04\,meV) is slightly larger than that of the $\beta$ band (0.99\,meV) (Fig. \ref{alphabeta}g).\\

\begin{figure*}[tbp]
\begin{center}
\includegraphics[width=1\textwidth,angle=0]{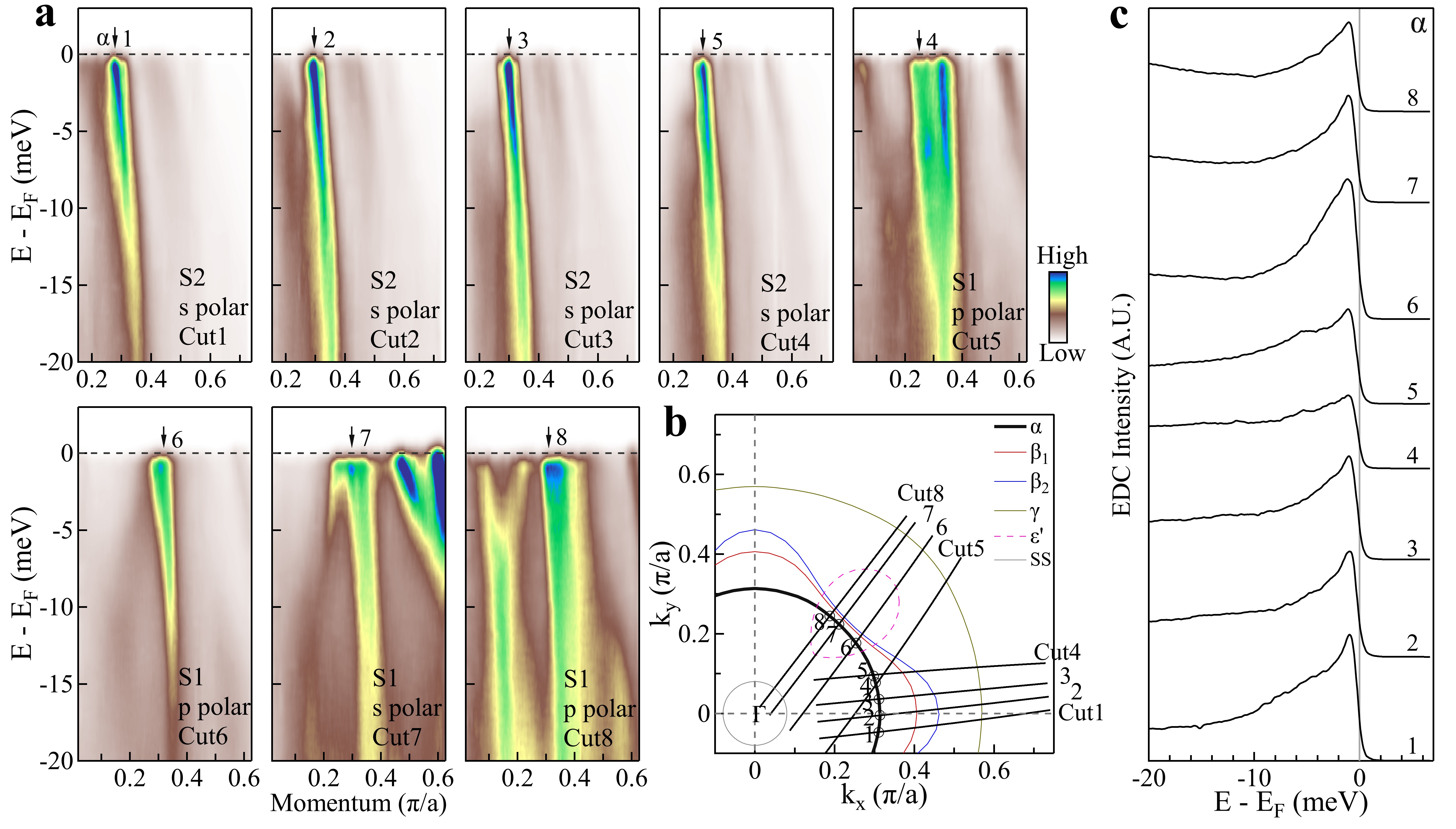}
\end{center}
\caption{{\bf EDCs along the $\alpha$ Fermi surface sheet measured at 0.9\,K in the superconducting state.} \textbf{a}, Band structures measured along different momentum cuts under different light polarizations on different samples. The location of the momentum cuts (Cut1-Cut8) are marked by black lines in \textbf{b}. Two light polarizations (s polarization and p polarization) are used and two samples (S1 and S2) are measured. The Fermi momentum location of the $\alpha$ band in the measured band structures is marked by arrows and labeled by numbers (1-8). \textbf{b}, Fermi surface of KFe$_2$As$_2$ marked with the momentum cuts (Cut1-Cut8) and the Fermi momentum points (1-8) along the $\alpha$ Fermi surface.
 \textbf{c}, EDCs along the $\alpha$ Fermi surface at the Fermi momentum points (1-8) extracted from \textbf{a}.
   }
\label{alpha}
\end{figure*}

\begin{figure*}[tbp]
\begin{center}
\includegraphics[width=1\textwidth,angle=0]{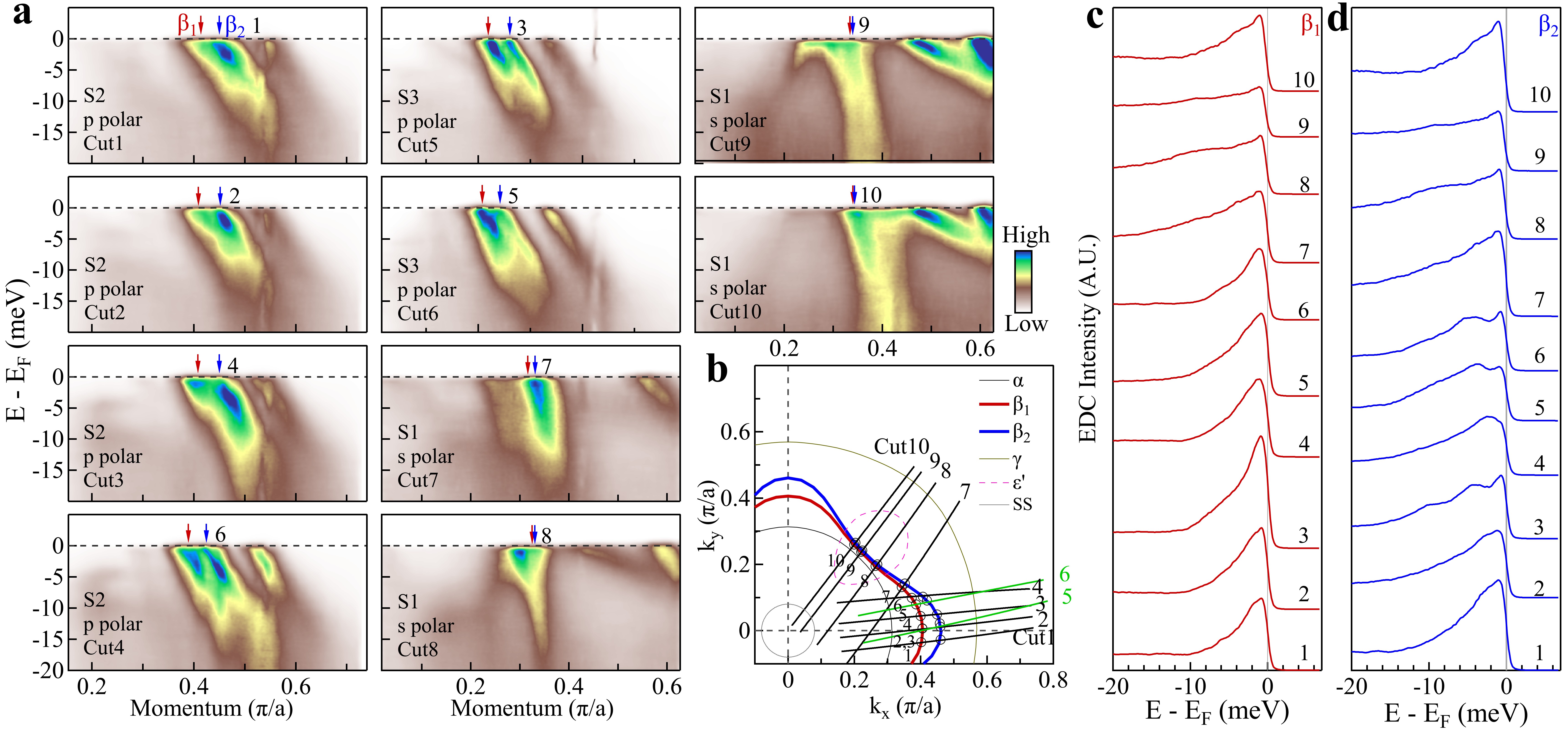}
\end{center}
\caption{{\bf EDCs along the $\beta_1$ and $\beta_2$ Fermi surface sheets measured at 0.9\,K in the superconducting state.} \textbf{a}, Band structures measured along different momentum cuts under different light polarizations on different samples. The location of the momentum cuts (Cut1-Cut10) are marked by black lines (1-4 and 7-10) and green lines (5-6) in \textbf{b}. Two light polarizations (s polarization and p polarization) are used and three samples (S1, S2 and S3) are measured. The Fermi momentum location of the $\beta_1$ and $\beta_2$ bands in the measured band structures is marked by red arrows and blue arrows, respectively. \textbf{b}, Fermi surface of KFe$_2$As$_2$ marked with the momentum cuts (Cut1-Cut10) and the Fermi momentum points (1-10) along the $\beta_1$ and $\beta_2$ Fermi surface sheets. \textbf{c}, EDCs along the $\beta_1$ Fermi surface at the Fermi momentum points (1-10) extracted from \textbf{a}. \textbf{d}, EDCs along the $\beta_2$ Fermi surface at the Fermi momentum points (1-10) extracted from \textbf{a}.
   }
\label{beta12}
\end{figure*}

\begin{figure*}[tbp]
\begin{center}
\includegraphics[width=1\textwidth,angle=0]{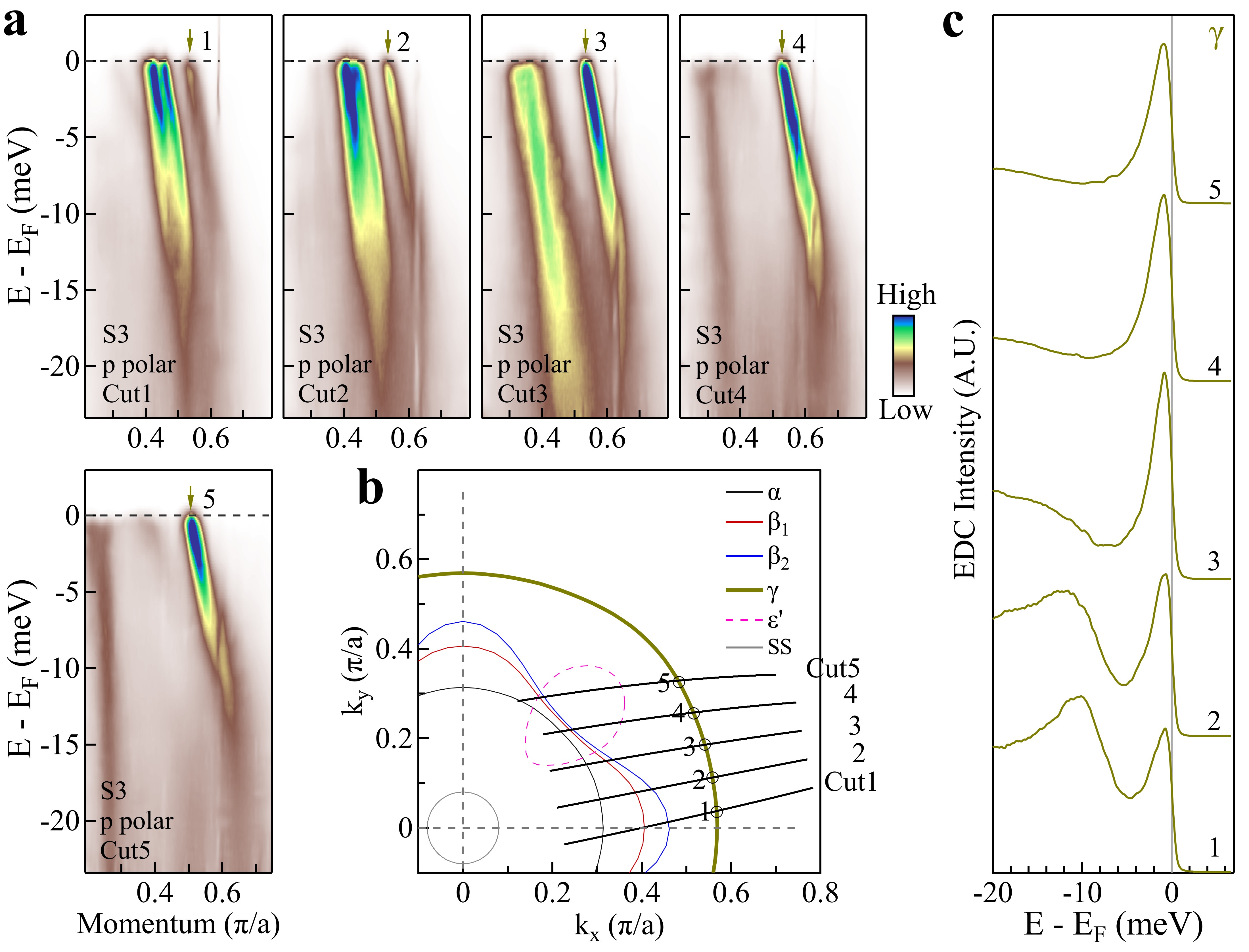}
\end{center}
\caption{{\bf EDCs along the $\gamma$ Fermi surface measured at 0.9\,K in the superconducting state.} \textbf{a}, Band structures measured along different momentum cuts. The location of the momentum cuts (Cut1-Cut5) are marked by black lines in \textbf{b}. The p polarization is used and the sample S3 is measured. The Fermi momentum location of the $\gamma$ band in the measured band structures is marked by arrows and labeled by numbers (1-5). \textbf{b}, Fermi surface of KFe$_2$As$_2$ marked with the momentum cuts (Cut1-Cut5) and the Fermi momentum points (1-5) along the $\gamma$ Fermi surface. \textbf{c}, EDCs along the $\gamma$ Fermi surface at the Fermi momentum points (1-5) extracted from \textbf{a}.
   }
\label{gamma}
\end{figure*}

\begin{figure*}[tbp]
\begin{center}
\includegraphics[width=1\textwidth,angle=0]{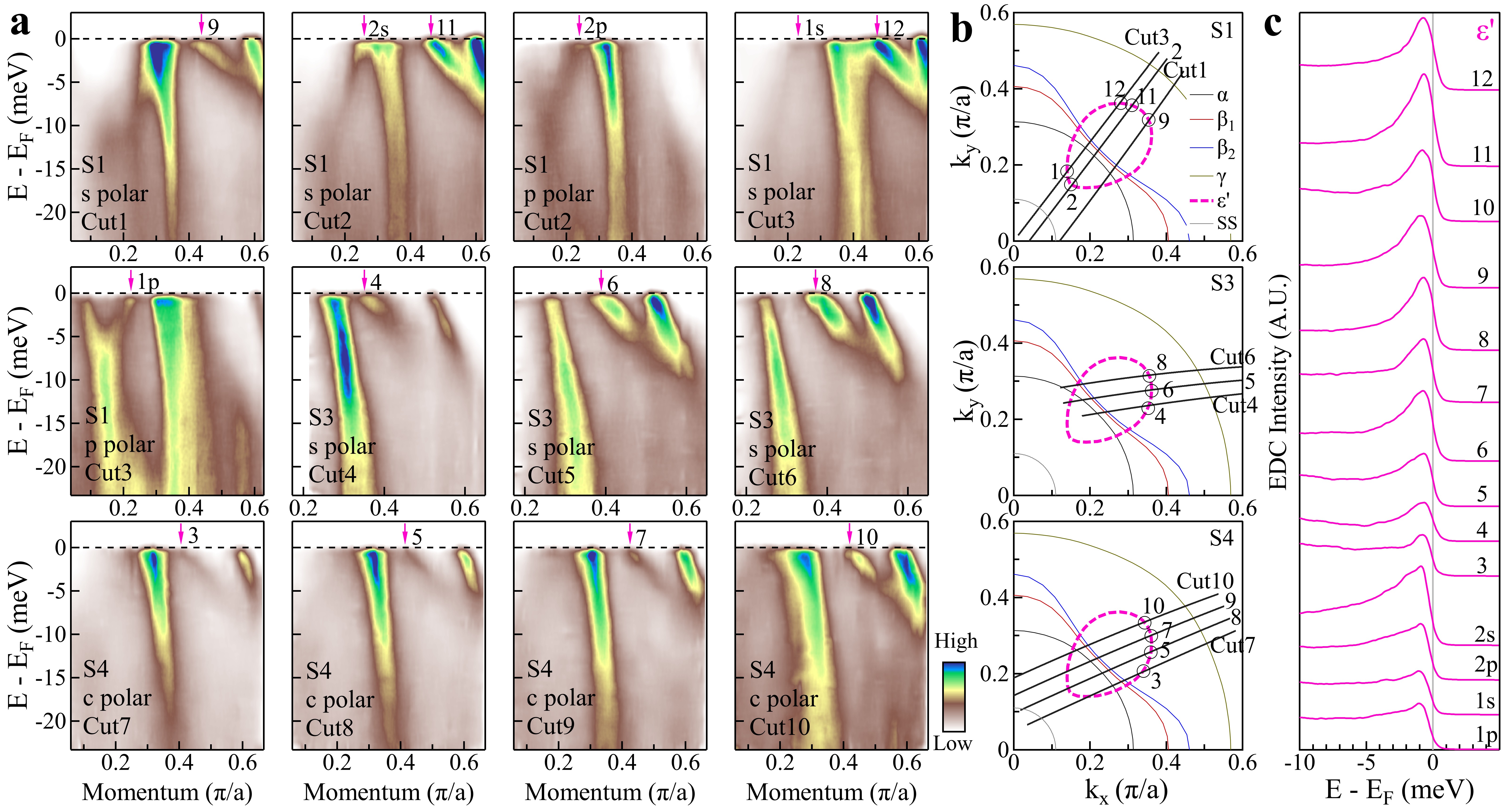}
\end{center}
\caption{{\bf EDCs along the $\varepsilon'$ Fermi surface measured at 0.9\,K in the superconducting state.} \textbf{a}, Band structures measured along different momentum cuts under different light polarizations on different samples. The location of the momentum cuts (Cut1-Cut10) are marked by black lines in \textbf{b}. Two light polarizations (s polarization and p polarization) are used and 3 samples (S1, S3 and S4) are measured. The Fermi momentum location of the $\varepsilon'$ band in the measured band structures is marked by arrows and labeled by numbers (1-12). \textbf{b}, Fermi surface of KFe$_2$As$_2$ marked with the momentum cuts (Cut1-Cut10) and the Fermi momentum points (1-12) along the $\varepsilon'$ Fermi surface. \textbf{c}, EDCs along the $\varepsilon'$ Fermi surface at the Fermi momentum points (1-12) extracted from \textbf{a}.
   }
\label{epsilon}
\end{figure*}

\begin{figure*}[tbp]
\begin{center}
\includegraphics[width=0.7\textwidth,angle=0]{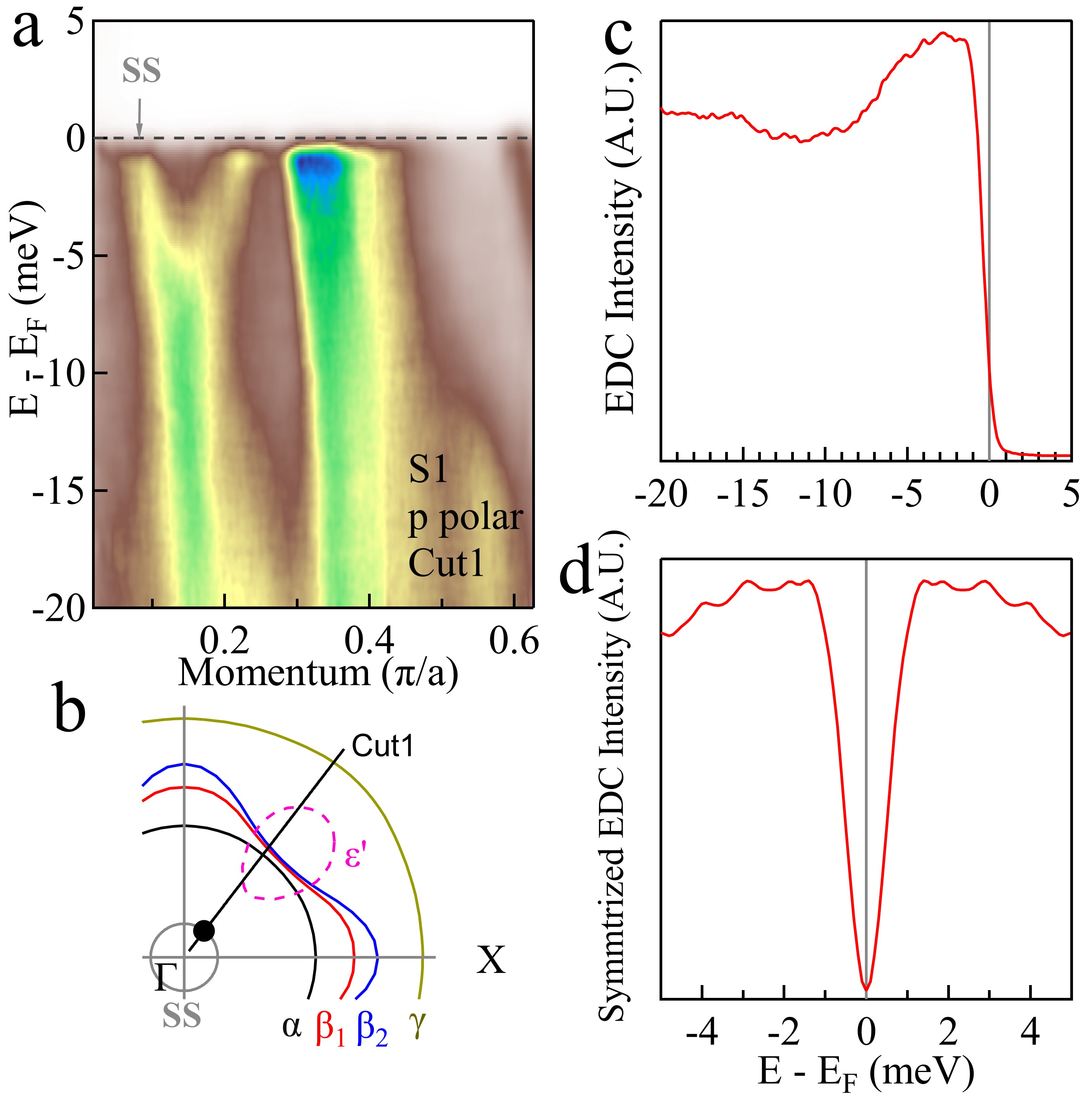}
\end{center}
\caption{{\bf An EDC along the surface state (SS) Fermi surface measured at 0.9\,K in the superconducting state.} \textbf{a}, A band structure measured along a momentum cut marked by the black line in \textbf{b}. The p polarization is used and the sample S1 is measured. The Fermi momentum location of the SS band in the measured band structure is marked by a grey arrow. \textbf{b}, Fermi surface of KFe$_2$As$_2$ marked with a momentum cut (Cut1) and the Fermi momentum point (the black circle) along the SS Fermi surface. \textbf{c}, An EDC along the SS Fermi surface at the Fermi momentum point extracted from \textbf{a}. \textbf{d}, Symmetrized EDC obtained from \textbf{c}.
   }
\label{SS}
\end{figure*}

\noindent\textbf{7. EDCs along Different Fermi Surface Sheets}\vspace{3mm}\\
\hspace*{6mm}In order to extract the superconducting gap along all the Fermi surface, we carried out ARPES measurements along different momentum cuts under different polarization geometries and repeated the measurements on many KFe$_2$As$_2$ samples. In Fig. \ref{alpha}-\ref{epsilon}, we show the measured band structures and the EDCs along the $\alpha$, $\beta$ ($\beta_1$ and $\beta_2$), $\gamma$ and $\varepsilon'$ Fermi surface, respectively. The EDC on the surface state Fermi surface around $\Gamma$ is shown in Fig. \ref{SS}.



\newpage
\clearpage

\vspace{3mm}
\vspace{3mm}
\newpage

\newpage

\newpage

\newpage
\thispagestyle{empty}

\clearpage
\begin{table}[ ]
\centering
\caption{\textbf{Fermi surface area, effective mass \bm{m$^*$} and Sommerfeld coefficient \bm{$\gamma$} of \bm{KFe$_2$As$_2$}. }}
\scalebox{0.6}{
\begin{tabular}{m{1.6cm}<{\centering} m{0.2cm}<{\centering}        m{1.9cm}<{\centering} m{1.7cm}<{\centering} m{1.4cm}<{\centering} m{2cm}<{\centering}m{0.4cm}<{\centering}                 m{3.3cm}<{\centering} m{1.9cm}<{\centering} m{1.8cm}<{\centering} m{1.8cm}<{\centering}m{0.5cm}<{\centering}           m{1.9cm}<{\centering}m{1.7cm}<{\centering} m{1.4cm}<{\centering} m{1.8cm}<{\centering}}
\hline\hline

&&\multicolumn{4}{c}{\textbf{Area ($\%$BZ)}}&         &\multicolumn{4}{c}{\textbf{\textbf{m}$^*$/m$_e$ (\textbf{m}$^*$/m$_b$)}}&               & \multicolumn{4}{c}{\bm{$\gamma$} \textbf{(mJ\,mol$^{-1}$\,K$^{-2}$)}}\\
\cline{3-6}\cline{8-11}\cline{13-16}

 \textbf{Fermi Surface}&         &\textbf{this work} &\textbf{previous ARPES}&\textbf{dHvA}&\textbf{Calculation}&         &\textbf{this work} &\textbf{previous ARPES}&\textbf{dHvA}&\textbf{Calculation}&                  &\textbf{this work} &\textbf{previous ARPES}&\textbf{dHvA}&\textbf{Calculation} \\
\hline
 $\alpha$&                  &7.6&9.5&8.4&21.2&                                         &6.1 (6.1) ($\Gamma$-M)&5.9 (4.3)&6.3 (4.6)&1.0($\Gamma$-M)&          &8.7&8.6&9.1&2.0\\
&                                     &&&&&                                                              &5.7 (3.2) ($\Gamma$-X)&&&1.8($\Gamma$-X)&                                            &&&& \\
$\beta$&          &11.0 ($\beta_1$)&14.6&13.0&13.0&           &8.0 (2.8) ($\beta_1$$\Gamma$-M)&14.4 (5.9)&13.3 (5.4)&2.8($\Gamma$-M)&    &15.5 ($\beta_1$)&20.9&19.2&3.6   \\
&                &&&&&                                                                        &13.3 (6.4) ($\beta_1$$\Gamma$-X)&&&2.1($\Gamma$-X)&                                   &&&&  \\
&                &13.2 ($\beta_2$)&&&&                                              & 8.0 (2.8) ($\beta_2$$\Gamma$-M)&&&&                                       &17.3 ($\beta_2$)&&&  \\
&                &&&&&                                                                      & 15.6 (7.5) ($\beta_2$$\Gamma$-X)&&&&                                         &&&&  \\

$\gamma$&              &26.1&28.7&25.6&17.1&                          &25.7 (6.1) ($\Gamma$-M)&17.1 (5.7)&19 (6.3)&4.2($\Gamma$-M)&                        &34.9&24.8&27.6&4.3 \\
&                                     &&&&&                                                    &22.4 (13.0) ($\Gamma$-X)&&&1.7($\Gamma$-X)&                                                               &&&& \\
$\varepsilon$ &       &0.97$\times$4&2.1$\times$4&1.1$\times$4&0.24$\times$4&        &13.8 (8.1) ($\varepsilon_R$)&4.9 (4.9)&6.6 (6.6)&1.4($\Gamma$-M)&         &12.4$\times$4&7.1$\times$4&9.6$\times$4&1.4$\times$4  \\
        &                    &&&&&                                                        &3.2 (16) ($\varepsilon_L$)&&&0.2($\Gamma$-X)&                                                            &&&&          \\

total&                              &49.7&61.2&51.4&51.5&                                &&&&&                                     &109.6&82.7&94&15.5\\\hline
&\\
\end{tabular}}
\raggedright
{\scriptsize{The Fermi surface area is extracted from the measured Fermi surface in Fig. 1b. The effective mass $m^*$ is extracted from $m^* =  {\hbar}k_F/v_F$. The Sommerfeld coefficient $\gamma$ is calculated from $\gamma$ = 1.452\,$\mu_0^ *$\,(mJ\,mol$^{-1}$\,K$^{-2}$)\cite{SUji2013TTerashima}, where $\mu_0^*$ = m$^*$/m$_e$, by neglecting the k$_z$ dispersion. The results from the previous ARPES measurements\cite{HHisatomo2014TYoshida}, dHvA measurements\cite{SUji2013TTerashima}, previous band structure calculations\cite{UShinya2010TTaichi,HHisatomo2014TYoshida} and our band structure calculations are also included.}}
\label{effeMass2}
\end{table}